 \documentclass[twocolumn,times]{aastex631}
\usepackage{amsmath,amssymb}

\def\mJyb{mJy beam$^{-1}$}
\newcommand{\obj}{\mbox{III~Zw~2}}
\newcommand{\kms}{km\,s$^{-1}$} 
\usepackage{color}
\newcommand{\angstrom}{\text{\normalfont\AA}}

\begin{document}

\title{Interactions between the jet and disk wind in a nearby radio intermediate quasar III Zw 2}
\shorttitle{Jet-wind interactions in III Zw 2}
\shortauthors{Wang et al.}

\correspondingauthor{Tao An}
\email{antao@shao.ac.cn}

\author[0000-0002-7351-5801]{Ailing Wang}
\affiliation{Shanghai Astronomical Observatory, CAS, Nandan Road 80, Shanghai, 200030, China; }
\affiliation{School of Astronomy and Space Sciences, University of Chinese Academy of Sciences, No. 19A Yuquan Road, Beijing 100049, China}

\author[0000-0003-4341-0029]{Tao An}
\affiliation{Shanghai Astronomical Observatory, CAS, Nandan Road 80, Shanghai, 200030, China; }
\affiliation{School of Astronomy and Space Sciences, University of Chinese Academy of Sciences, No. 19A Yuquan Road, Beijing 100049, China}

\author[0000-0003-0181-7656]{Shaoguang Guo}
\affiliation{Shanghai Astronomical Observatory, CAS, Nandan Road 80, Shanghai, 200030, China; }
\affiliation{School of Astronomy and Space Sciences, University of Chinese Academy of Sciences, No. 19A Yuquan Road, Beijing 100049, China}

\author[0000-0002-2211-0660]{Prashanth Mohan}
\affiliation{Shanghai Astronomical Observatory, CAS, Nandan Road 80, Shanghai, 200030, China; }

\author[0000-0001-7350-4152]{Wara Chamani}
\affiliation{Aalto University Mets\"ahovi Radio Observatory, Mets\"ahovintie 114, 02540 Kylm\"al\"a, Finland}
\affiliation{Aalto University Department of Electronics and Nanoengineering, P.O. BOX 15500, FI-00076 AALTO, Finland}

\author[0000-0003-3389-6838]{Willem A. Baan}
\affiliation{Xinjiang Astronomical Observatory, Chinese Academy of Sciences, 150 Science 1-Street, 830011 Urumqi, P.R. China;}
\affiliation{Netherlands Institute for Radio Astronomy ASTRON, NL-7991 PD Dwingeloo, the Netherlands;}

\author{Talvikki Hovatta}
\affiliation{Finnish Centre for Astronomy with ESO, University of Turku, Vesilinnantie 5, FI-20014, Finland}
\affiliation{Aalto University Mets\"ahovi Radio Observatory, Mets\"ahovintie 114,
02540 Kylm\"al\"a, Finland}

\author{Heino Falcke}
\affiliation{Department of AstrophysicsInstitute for Mathematics, Astrophysics and Particle Physics, Radboud University Nijmegen, PO Box 9010, 6500 GL, Nijmegen, The Netherlands}

\author[0000-0002-2801-766X]{Tim J. Galvin}
\affiliation{CSIRO Space and Astronomy, PO Box 1130, Bentley, WA 6102, Australia}
\affiliation{International Centre for Radio Astronomy Research, Curtin University, Bentley, WA 6102, Australia}

\author[0000-0002-5119-4808]{Natasha Hurley-Walker}
\affiliation{International Centre for Radio Astronomy Research, Curtin University, Bentley, WA 6102, Australia}

\author[0000-0002-5125-695X]{Sumit Jaiswal}
\affiliation{Shanghai Astronomical Observatory, CAS, Nandan Road 80, Shanghai, 200030, China; }

\author[0000-0002-0393-0647]{Anne Lahteenmaki}
\affiliation{Aalto University Mets\"ahovi Radio Observatory, Mets\"ahovintie 114,
02540 Kylm\"al\"a, Finland}
\affiliation{Aalto University Department of Electronics and Nanoengineering, P.O. BOX 15500, FI-00076 AALTO, Finland}

\author[0000-0002-3426-3269]{Baoqiang Lao}
\affiliation{Shanghai Astronomical Observatory, CAS, Nandan Road 80, Shanghai, 200030, China; }
\affiliation{School of Physics and Astronomy, Yunnan University, Kunming, 650091, China }

\author{Weijia Lv}
\affiliation{Shanghai Astronomical Observatory, CAS, Nandan Road 80, Shanghai, 200030, China; }

\author[0000-0003-1249-6026]{Merja Tornikoski}
\affiliation{Aalto University Mets\"ahovi Radio Observatory, Mets\"ahovintie 114,
02540 Kylm\"al\"a, Finland}

\author[0000-0001-8256-8887]{Yingkang Zhang}
\affiliation{Shanghai Astronomical Observatory, CAS, Nandan Road 80, Shanghai, 200030, China; }



\begin{abstract}
Disk winds and jets are ubiquitous in active galactic nuclei (AGN), and how these two components interact remains an open question. We study the radio properties of a radio-intermediate quasar III Zw 2. We detect two jet knots J1 and J2 on parsec scales, which move at a mildly apparent superluminal speed of $1.35\,c$. Two $\gamma$-ray flares were detected in III Zw 2 in 2009--2010, corresponding to the primary radio flare in late 2009 and the secondary radio flare in early 2010. The primary 2009 flare was found to be associated with the ejection of J2. The secondary 2010 flare occurred at a distance of $\sim$0.3 parsec from the central engine, probably resulting from the collision of the jet with the accretion disk wind. The variability characteristics of III Zw 2 (periodic radio flares, unstable periodicity, multiple quasi-periodic signals and possible harmonic relations between them) can be explained by the global instabilities of the accretion disk. These instabilities originating from the outer part of the warped disk propagate inwards and can lead to modulation of the accretion rate and consequent jet ejection. At the same time, the wobbling of the outer disk may also lead to oscillations of the boundary between the disk wind and the jet tunnel, resulting in changes in the jet-wind collision site. III Zw 2 is one of the few cases observed with jet-wind interactions, and the study in this paper is of general interest for gaining insight into the dynamic processes in the nuclear regions of AGN.

\end{abstract}

\keywords{galaxies (563) --- Active galactic nuclei (16) --- Very long baseline interferometry (1769) --- Radio jets (1347)}


\section{Introduction}
Disk winds and jets are ubiquitous in active galactic nuclei (AGN) \citep{2014ARA&A..52..529Y,2019ARA&A..57..467B} and play an important role in the AGN feedback to their host galaxies \citep{2015ARA&A..53..115K,2018NatAs...2..198H,2021NatAs...5..928S}.  
Radio-loud (RL) AGN  \footnote{AGNs are divided into radio-loud (RL AGN) and radio-quiet AGN (RQ AGN) by the flux density ratio of radio (5 GHz) to optical ($\rm 4400 \angstrom$) continuum, the so-called radio-loudness parameter ($R=S_{\rm 5 GHz}/S_{\rm 4400 \angstrom}$)  \citep{1989AJ.....98.1195K}. RL AGN have $R\geq10$ and RQ AGN have $R<10$. } comprise about 10\% of the entire AGN population \citep[e.g. ][]{2002AJ....124.2364I}; their radio emission is dominated by relativistic jets  \citep{1995PASP..107..803U,2019ARA&A..57..467B}. The energy release from radio-quiet (RQ) AGN, which occupy the majority of the AGN population, is dominated by thermal emission related to the accretion disk  \citep{2016A&ARv..24...13P,2019NatAs...3..387P}. Observational evidence and magnetohydrodynamic models suggest that low-power jets and winds may coexist in RQ AGN \citep[e.g. ][]{2012MNRAS.424..754T,2014ApJ...780..120F,2017A&A...600A..87G}. However, whether and how the jet and disk wind interact with each other remains an open question  \citep{2019NatAs...3..387P}. 
Some studies suggest that there exists a class of objects with moderate radio loudness, called radio-intermediate (RI) AGN \citep{1993MNRAS.263..425M}. Observing RI AGN is much less difficult than RQ AGN, and RI AGN have mixed properties of RL and RQ AGN, thus providing an opportunity to study jet- and wind-driven AGN feedback and jet-wind interactions.

\obj\ (\citealt{1967AdA&A...5..267Z}, also named PG~0007+106, Mrk~1501) is an unusual AGN containing many enigmatic observational characteristics.  It is hosted by a spiral galaxy \citep{1982ApJ...262...48H,1983Natur.303..584H} at redshift $z = 0.0893$  \citep{1970ApJ...160..405S}, with a prominent tidal arm \citep{2001AJ....122.2791S} to the north of the nucleus, but showing spectroscopic characteristics of a typical type I Seyfert  galaxy \citep{1968ApJ...152.1101A,1977ApJ...215..733O}. It is further included in the bright quasar sample \citep{1983ApJ...269..352S} with a bolometric luminosity up to $\approx 10^{45}$ erg s$^{-1}$ \citep{1982ApJ...253..485P,1995A&A...298..375F}. 

In radio bands, \obj\ is identified as a RI AGN \citep{1996ApJ...471..106F} with a radio loudness of $\sim$200 \citep{1989AJ.....98.1195K,1994AJ....108.1163K}. The source shows a large extended structure on kilo-parsec (kpc) scales \citep{1987MNRAS.228..521U,2005A&A...435..497B}, but its radio emission on parsec (pc) scales shows blazar-like behavior \citep{1999ApJ...514L..17F,2016ApJS..226...17L}. The Very Large Array (VLA) images of \obj\ show a triple radio structure extending along the northeast-southwest (NE--SW) direction with a total extent over 36$^{\prime\prime}$ \citep{2005A&A...435..497B}. The SW lobe is brighter but shorter than the NE lobe. The recently published images of \obj\ observed by the upgraded Giant Metrewave Radio Telescope (uGMRT) at 685 MHz  \citep{2020MNRAS.499.5826S,2021MNRAS.507..991S} show additional faint structures beyond the SW and NE lobes previously detected by the VLA, and these outer lobes extend along the direction perpendicular to the NE--SW jet. 

The most attractive feature of \obj\ is its extreme variability: over 20 fold in radio  \citep{1977IAUC.3145....2W,1978ApJ...222L..91S,1985ApJS...59..513A,1992A&AS...94..121T,1999ApJ...514L..17F,2005A&A...435..497B} and 10-fold in the X-ray band \citep{1988A&A...198...16K,2002PASA...19...73S,2002MNRAS.335..177S}; also highly variable in optical \citep{1980Natur.284..410L,1987MNRAS.226..137S}, infrared  \citep{1984MNRAS.209..697L,1995AJ....110..529C} and gamma-ray \citep{2016ApJS..226...17L} bands. Moreover, the large flares of \obj\ in multiple bands are found to be correlated \citep{2002PASA...19...73S}. 
The variability characteristics of \obj\ is very rare in RI and RQ AGN, and instead is similar to blazars \citep{2003ApJ...586...33A,2004A&A...427..769T,2011ApJS..194...29R}, making it stand out from the RI and RQ AGN populations.  In addition, the radio flares seem to exhibit quasi-periodic variations with a period of about 4--5 years \citep{2003PASA...20..126B,2010NewA...15..254L}, which warrants an in-depth study of the physical mechanisms underlying the periodic variability.
 
Studying the structural changes and variability of the jet during prominent flare phases can help reveal the mechanism of jet production and evolution. Imaging the newly born jet features requires sub-parsec resolutions, which is currently only possible with Very Long Baseline Interferometry (VLBI) observations. In the past four decades, several large radio flares have been found in \obj\  \citep{1985ApJS...59..513A,1992A&AS...94..121T,2005A&A...435..497B,2010NewA...15..254L}, with flux density approaching or exceeding 3 Jy at 15 GHz.  VLBI observations during the 1998 flare revealed a compact core-jet structure within 0.1--0.4 parsec \citep{1996ApJ...473L..13F,1998AJ....115.1295K,1999ApJ...514L..17F,2000A&A...357L..45B}. 
A dramatic structural change was found from the 43 GHz VLBI data between 1998 December 12 and 1999 July 15, and an apparent superluminal speed of $1.25\,c$ \citep{2000A&A...357L..45B} was inferred, making \obj\ the first RI AGN detected with apparent superluminal jet motion in a spiral galactic nucleus  \citep{2000A&A...357L..45B,2005A&A...435..497B}. After the major flare in 2009, the variability amplitude became smaller (see Figure 4 in  \citealt{2005A&A...435..497B}). In the latest VLBI observations of \obj\ in 2017, only one compact core was detected and the core was significantly fainter than $\sim20$ years ago  \citep{2021A&A...652A..14C}.

In this paper, we investigate the jet kinematics and radio variability and propose a model to explain the quasi-periodic variability, the correlation between the prominent flare and jet production, and the secondary flare created by the jet-wind collision.

\section{Data} \label{sec:data}

The discovery of the southern extended feature from uGMRT images \citep{2020MNRAS.499.5826S,2021MNRAS.507..991S} motivated us to use low-frequency interferometric data, including the data from the Australian Square Kilometre Array Pathfinder (ASKAP, \citealt{2021PASA...38....9H}) at 888 MHz and GMRT \citep{1991CSci...60...95S}  at 150 MHz (the TIFR GMRT Sky Survey, TGSS, \citealt{2017A&A...598A..78I}) and the Murchison Widefield Array (MWA, \citealt{2013PASA...30....7T}) observations at 72--231 MHz  (Appendix \ref{app:MWAdata}) to study the complete radio structure, to reveal the kpc-scale morphology, and to constrain the radio spectrum of the extended structure.
The VLBI data used for the jet kinematics analysis are obtained from the Monitoring Of Jets in Active galactic nuclei with VLBA Experiments (MOJAVE, \citealt{2018ApJS..234...12L}) program and the archived data from the Astrogeo database \footnote{\url{http://astrogeo.org/}}. Details of the VLBI data reduction are referred to Appendix \ref{app:VLBIdata}. The radio light curve data are obtained from the single-dish monitoring programs of the Owens Valley Radio Observatory at 15 GHz and the Mets\"{a}hovi radio telescope at 37 and 22 GHz (Appendix \ref{app:singledish}).

\section{Results and discussion} \label{sec:results}

\subsection{The jet structure}\label{sec:S-shape}

The VLA images of \obj\ in the literature show a triple structure along the NE--SW direction \citep{2005A&A...435..497B}. The central core C dominates the total flux density at almost all wavelengths.
The SW lobe is located at $15\farcs4$ ($\sim$25.5 kpc) \citep{1987MNRAS.228..521U,1998MNRAS.297..366K,1999ApJ...514L..17F} and is connected to the central core by a curved jet \citep{2020MNRAS.499.5826S}: the jet initially points to the west and then gradually bends to the southwest. A weaker lobe is detected at $21\farcs9$ ($\sim$36 kpc) northeast of the core \citep{2005A&A...435..497B}.

\begin{figure}
    \centering
    \includegraphics[width=0.45\textwidth]{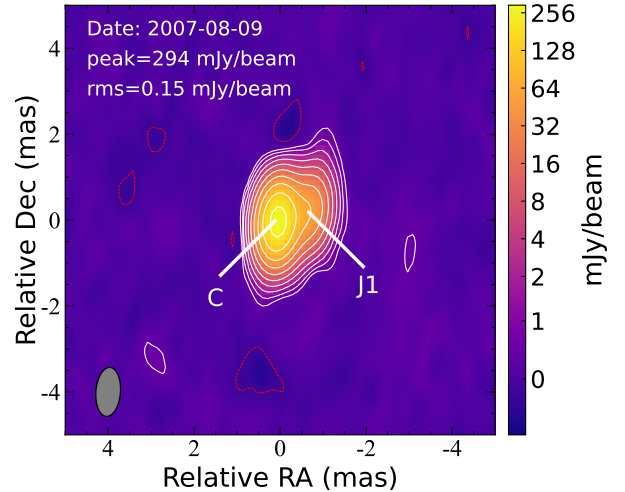}
    \caption{15-GHz VLBA image of \obj. The image is created with natural weighting. The negative red colored contour is at $-2\sigma$ level and the positive white-colored contours are in the series of 3$\sigma \times$(1, 2, 4, 8, 16, 32, 64, 128, 256, 512), where the \textit{rms} noise is $\sigma = 0.15$ \mJyb{}. The color scale shows the intensity in the logarithmic scale. }
    \label{fig:vlbimap}
\end{figure}

The VLBI image in Figure~\ref{fig:vlbimap} shows that the pc-scale jet extends to the west and is roughly aligned with the SW lobe (Figure~\ref{fig:mor}); therefore, it is possible that the SW jet is at the advancing side. If the SW and NE lobes were formed from the AGN activity in the same episode, then the length of the advancing jet/lobe should be longer than the reverse jet/lobe. However, the actual situation is the opposite. Although the SW lobe is brighter than the NE lobe, both the VLA and ASKAP images (Figure~\ref{fig:mor}) show that the SW lobe is shorter than the NE lobe. 
The difference in length between the two lobes seems to indicate that the ambient interstellar medium (ISM) on two opposite sides of the nucleus has different properties \citep{2005A&A...435..497B,2021MNRAS.507..991S}, i.e., the ISM at the southwest side has a higher density and the growth of the southwest jet is more obstructed by the ISM. A companion galaxy \citep{2001AJ....122.2791S} exists $\sim30\arcsec$ south of \obj\ (Figure~\ref{fig:tidalarm}), and gravitational interactions may have accumulated a large amount of gas in the intermediate region between the two galaxies.  The SW jet bends southward at the SW lobe, where enhanced polarization is observed \citep{2021MNRAS.507..991S}, providing observational evidence for strong interactions between the SW jet and the ISM. 

The VLBI image in Figure~\ref{fig:vlbimap} shows a compact jet with a maximum extent of 1.2 mas (corresponding to a projected distance of $\sim$2 pc). 
The integrated flux densities obtained from the VLBI images match those obtained from single-dish observations in close epochs, indicating that the contribution of the optically-thin extended jet to the total flux density at GHz frequencies is very small.
Another intrinsic reason for the compact jet is that the radio activity of \obj\ is intermittent and the VLBI jet is short-lived in nature (see discussion in Sections \ref{sec:kinematics} and \ref{sec:temporalevo}), which also lead to a short VLBI jet.
A similar situation can be seen in another RI AGN Mrk~231 \citep{2017ApJ...836..155R,2021MNRAS.504.3823W}.

Combining images of \obj\ observed with different resolutions at different frequencies, we find that the jet exhibits an overall S shape.
There are many physical mechanisms that can create S- or Z-shaped jet morphology, including: 
backflow from hotspots   \citep{1984MNRAS.210..929L},  buoyancy force bending the outer radio structure into the direction of decreasing external gas pressure \citep{1995ApJ...449...93W}, 
or precession of the jet beam  \citep{1978Natur.276..588E}, 
or jet reorientation due to hydrodynamic processes associated with galaxy mergers \citep{ 2003ApJ...594L.103G}.
\obj\ lives in a cluster environment and exhibits ongoing galaxy merger or galaxy-galaxy interactions \citep{2001AJ....122.2791S}, with observational evidence of an optical tidal arm north of the \obj\ nucleus and a companion galaxy $\sim30^{\prime\prime}$ south of the nucleus (see Figure \ref{fig:tidalarm}).
The spin-flip of the central engine following the galaxy merger can lead to a re-orientation of the jet, which can produce the S- or X-shape jet \citep{2003ApJ...594L.103G,2007ApJ...661L.147B}.
Using a typical advance speed of the terminal hotspot (e.g., M87, $v=0.11\,c$, \citealt{1995ApJ...447..582B}) as a reference, we obtain a kinematic age of about $\sim10^6$~yr for the extended \obj\ lobe. The \obj\ jet is not as powerful as the M87 jet and may have a lower hotspot advance speed, which would result in an estimated kinematic age of more than 1 Myr, but should not be larger than $10^7$~yr. This kinematic age is within the timescale of the black hole coalescence \citep{1991Natur.354..212E} and the typical lifetime of an AGN, allowing for the jet re-orientation as a possible mechanism for the S-shaped morphology.

\subsection{Jet properties at the pc-scale}\label{sec:kinematics}

\begin{figure*}
    \centering
    \includegraphics[width=0.9\textwidth]{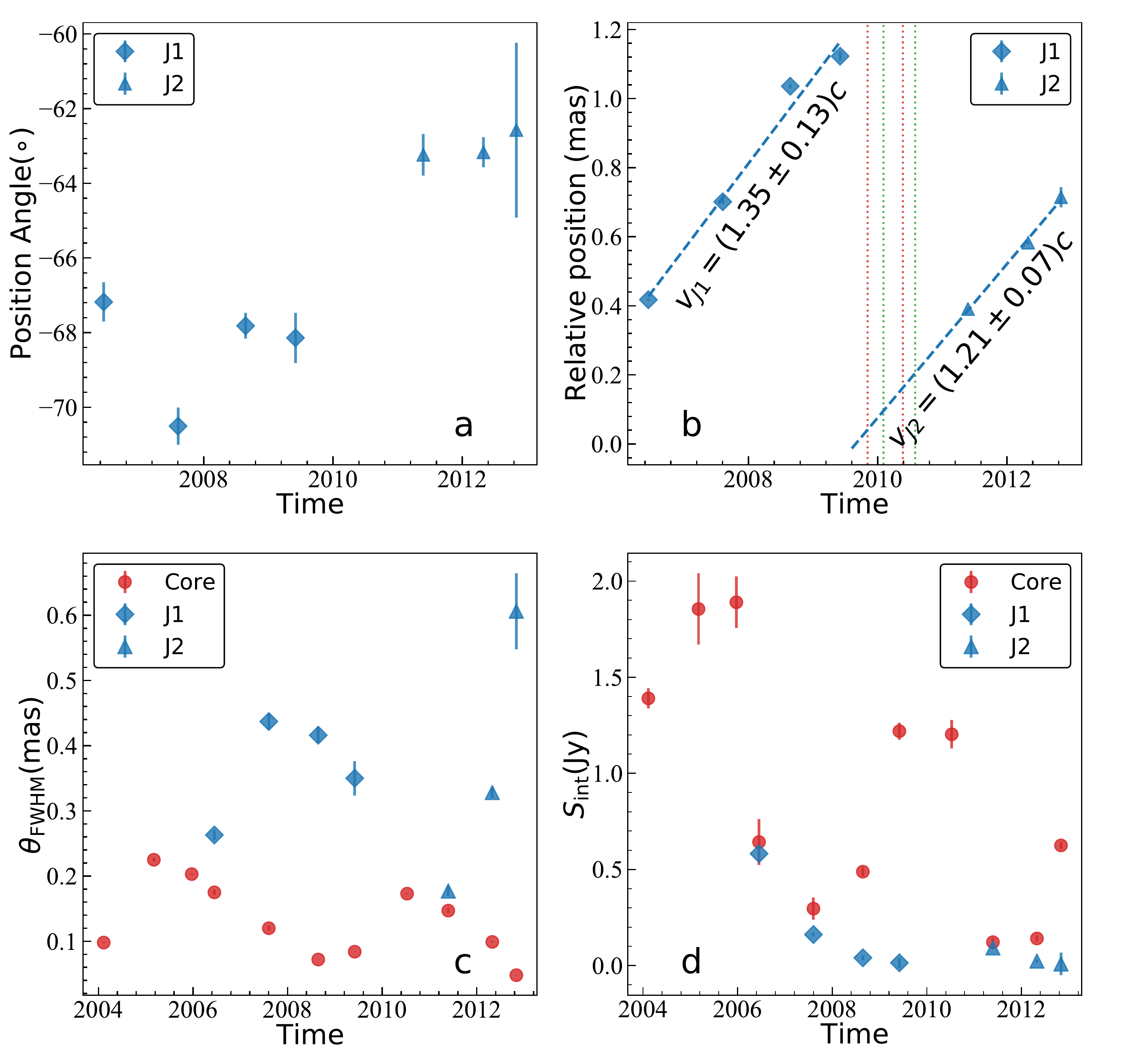}
    \caption{Properties of VLBI properties. a) changes of jet position angle with time; b) jet proper motion. The blue dashed lines represent the linear regress fit whose slope gives the jet proper motion; red dotted lines and green dotted lines remark the time of $\gamma$-ray flare peak and the time of radio flare peak at 15 GHz, respectively; c) changes of VLBI components' size with time; d) the changes of the integrated flux densities of VLBI components with time. To make the trend of the integrated flux density of the jets more visible, the integrated flux density of the jets has been multiplied by a factor of 2. The fitted parameters are referred to Table~\ref{table2}.  Extrapolating the linear fitting results in the jet ejection time, 2004 September and 2009 August for J1 and J2, respectively.}
    \label{fig:VLBI-2}
\end{figure*}

The 15-GHz MOJAVE archive data of \obj\ covers a time span of up to $\sim$18 years from July 1995 to June 2013, containing 25 epochs. The sub-mas resolution, high sensitivity and homogeneous image quality of the MOJAVE VLBI images make them ideally suited for studying jet kinematics. 
Although the jet is very compact, it can be well distinguished from the core in the VLBA images.
Figure \ref{fig:VLBI-2}-a shows the variation of the jet position angle with time and Figure \ref{fig:VLBI-2}-b shows the variation of the core-jet separation with time. It is clear that these jet knots belong to two separate components (labeled J1 and J2), rather than a single component. J1 and J2 follow their respective ballistic trajectories at different position angles.

We only used the jet components with the highest confidence, and the data in the other epochs were discarded because the data quality was too poor to obtain a reliable model fit (Appendix \ref{app:VLBIdata}).
We obtain  proper motion velocities of $v (J1) = 1.35 \pm 0.13 \,c$ and $v (J2) = 1.21 \pm 0.07 \,c$, consistent with those obtained by the MOJAVE team using all fitted components \citep{2019ApJ...874...43L}, and also in good agreement with previous studies based on the 43-GHz VLBI observations \citep{2000A&A...357L..45B}.

\begin{figure*}
    \centering
    \includegraphics[width=1\textwidth]{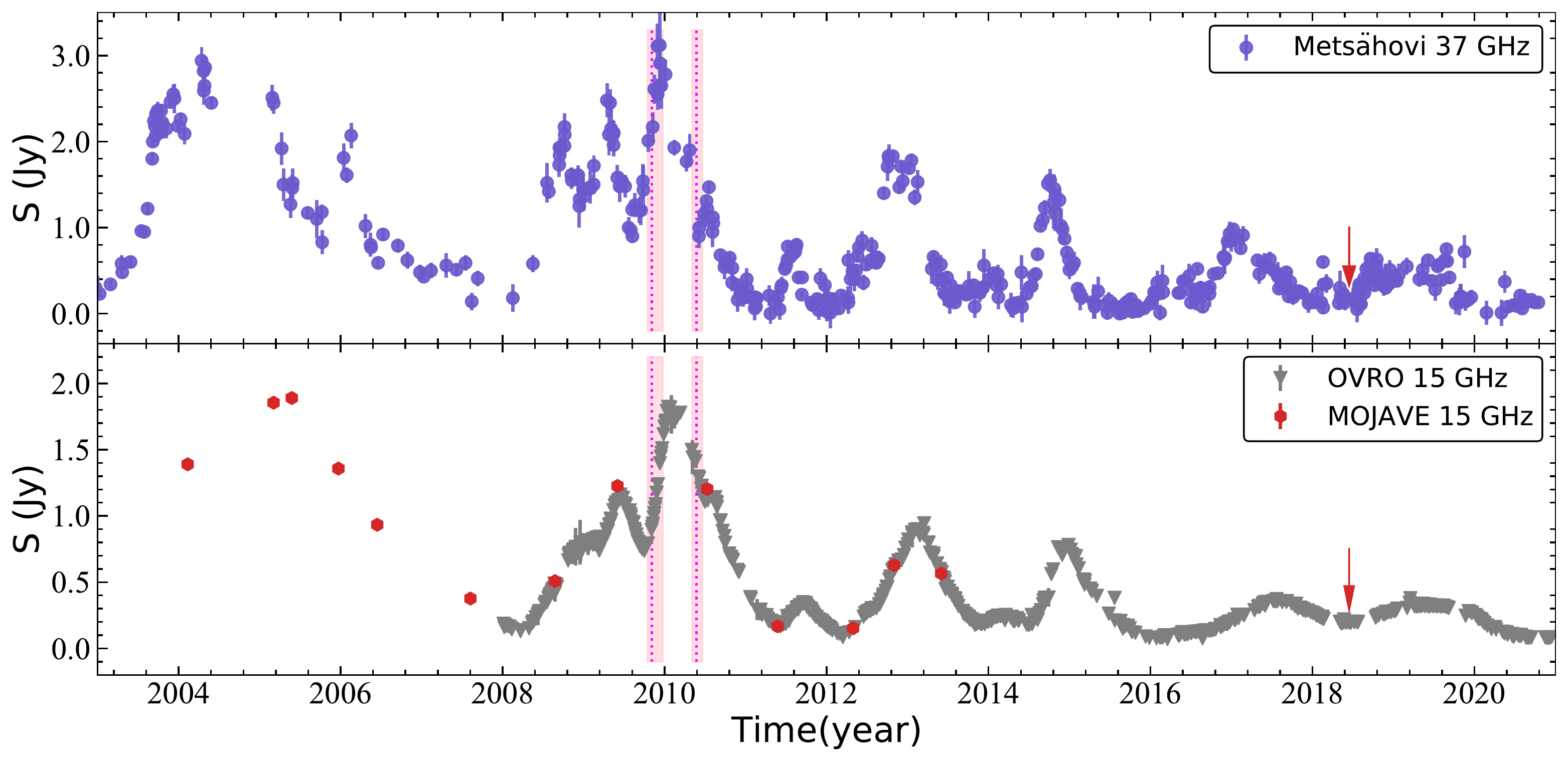}
    \caption{Radio light curves of \obj.  The 15-GHz data are observed with the 40-meter telescope of the Owens Valley Radio Observatory (OVRO). The 37 GHz data are observed with 14-meter radio telescope of the Mets\"{a}hovi Radio Observatory (MRO). The MRO data from 1986 to 2019 have been published in \citet{2020A&A...635A.172C} and other previous studies; in addition to these data, we include the new data from 2019 onwards to date. The VLBI data are from the MOJAVE program \citep{2018ApJS..234...12L}. The arrow marks the time of the MWA observation \citep{2022PASA...39...35H}. The red vertical shaded areas mark the time of $\gamma$-ray flare occurrence and the dashed lines correspond to the peaks of the two $\gamma$-ray flares \citep{2016ApJS..226...17L}. }
    \label{fig:lightcurve}
\end{figure*}

Assuming that the velocity of J1 remains constant from the time it was created until it disappeared, the ejection time of J1 can be traced back to epoch 2004.74 and is temporally correlated with the 2004 flare. Similarly, the ejection of J2 is associated with the 2009 flare. 
From 1996 onwards, three major radio flares over 3 Jy were observed at 37 GHz, with their peaks in early 1999 \citep{2005A&A...435..497B}, late 2004 \citep{2020A&A...635A.172C} and 2009 (Figure \ref{fig:lightcurve} 
in the present paper).
Each flare is associated with the creation of a discrete jet knot (the first discovered superluminal jet knot in \citealt{2000A&A...357L..45B} and J1 and J2 reported in this paper), possibly connected with intermittently enhanced accretion  (see more discussion in Section \ref{sec:variability}). 

Some lower-level flares that occurred during the intervals between these major flares, such as those in 2003 and 2006,  did not produce identifiable long-lived jet knots.  
Although the lack of long-term follow-up monitoring of the 1998 jet prevents us from obtaining information on how it evolved over time, our observations clearly indicate that the proper motion speeds of J1 and J2 are almost the same as that of the 1998 jet, suggesting that the fundamental condition for the jet production has not changed significantly during these jet creation events.

The core brightness temperatures  (column 7 of Table \ref{table2}) calculated from the VLBI-based measurements (core flux density and size) all exceed the equipartition brightness temperature limit \citep{1994ApJ...426...51R}. 
The Doppler boosting factor can be estimated as: $\delta = T_{\rm b,obs}/ T_{\rm b,eq}$, and the Lorentz factor $\Gamma$ can be estimated by $\Gamma = (\delta^2 + \beta_{\rm app}^2 + 1)/(2\delta)$. We find that large $\Gamma$ values are related to the major flares.  
In the quiescent state, we get a mean value of $\delta = 2.6$. Taking into account the jet proper motion speed $\beta_{\rm app} = 1.35\,c$, that yields a viewing angle of approximately $20^\circ$, which is a typical value for a non-blazar radio quasar \citep{1993ApJ...407...65G}. A larger viewing angle of $35\fdg4$ was derived by \citet{2009A&A...494..527H} and an upper limit of $41^\circ$ by \citet{2000A&A...357L..45B} due to different Doppler factor and jet speed used.

We calculated the magnetic field strength at 1 pc to be $\sim 53$ mG (Appendix \ref{app:emissionproperties}), similar to that estimated based on the apparent core shift and synchrotron self-absorption \citep{2021A&A...652A..14C}. This consistency supports the low jet magnetic flux in \obj\ and the inference that the central engine did not reach the magnetically arrested disk state \citep{2021A&A...652A..14C}. Interestingly, our estimate is based on data from the flaring phases while the earlier estimates by Chamani et al. are based on observations in a quiescent phase, suggesting that the jet structure remains relatively unperturbed by any fresh injection events at the jet base (at a distance of $\sim$ 800 gravitational radii).

\subsection{Radio spectrum} 
\label{sec:spectrum}

\begin{figure}
    \centering
    \includegraphics[width=0.9\columnwidth]{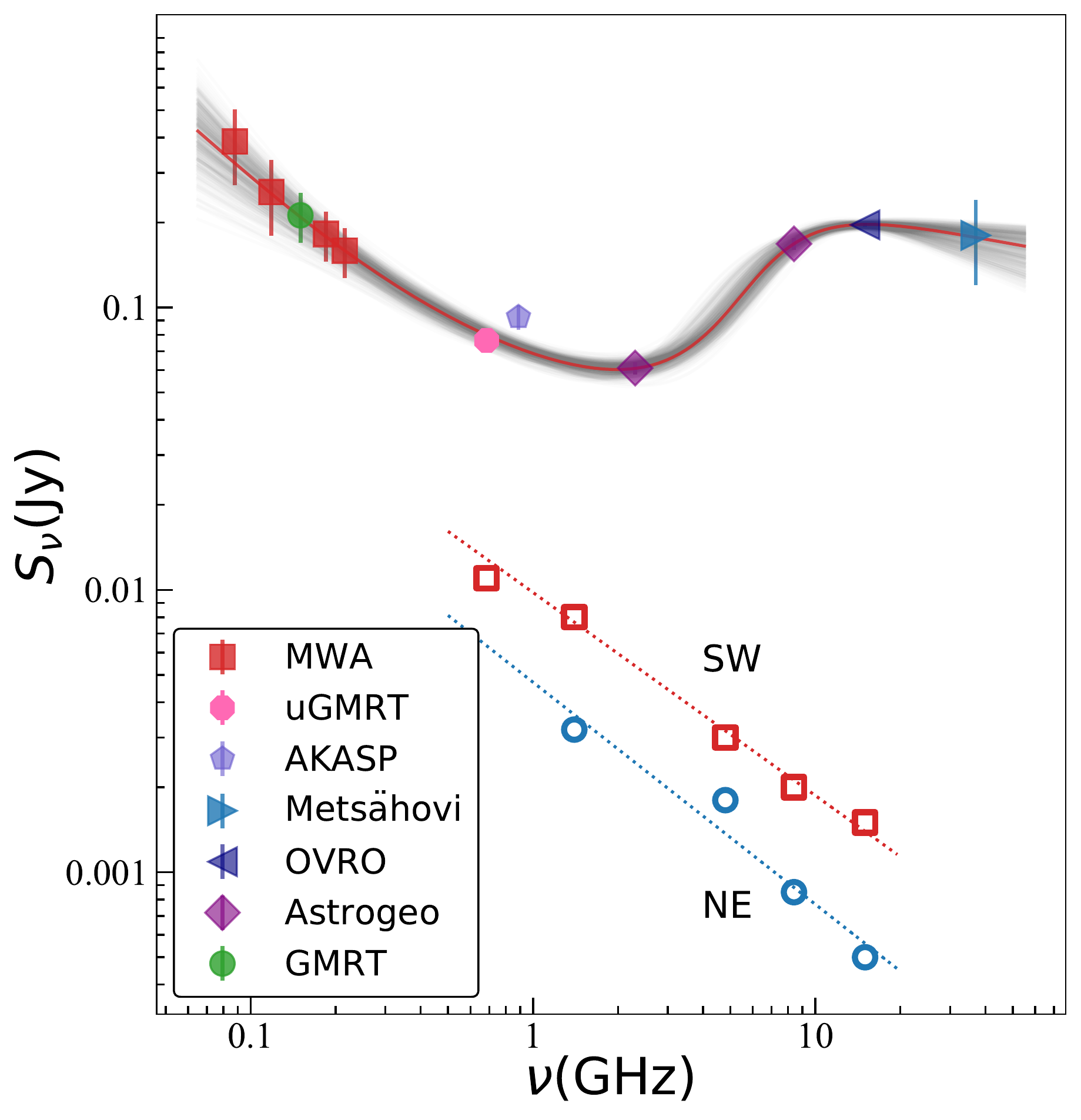}
    \caption{Radio spectrum of \obj. The red-colored squares are taken from the MWA GLEAM-X survey. The green-colored solid cycle represents the GMRT 150-MHz data point \citep{2017A&A...598A..78I}. The magenta-colored data point is the total flux density from the uGMRT observation \citep{2020MNRAS.499.5826S} on 2018 November 23. The slateblue data point is the total flux density from the ASKAP observation \citep{2020PASA...37...48M} on 2020 May 3. The blue-colored,  navy-colored and purple-colored data points are from this paper, respectively: 37 GHz Mets\"{a}hovi, 15 GHz OVRO, 8.4 and 2.3 GHz Astrogeo VLBI; for these data, we select the epochs close to the MWA and uGMRT observations in 2018. Core-dominated flux densities show an inverted spectrum with the best-fit self-absorbed synchrotron spectrum and more details are presented in Section \ref{sec:spectrum} and Appendix \ref{app:spectrum}. The red-colored open squares and blue-colored open circles are  the flux densities of the SW and NE lobes derived from the VLA images \citep{2005A&A...435..497B}.}
    \label{fig:SED}
\end{figure}

We plot the radio spectrum of \obj\ from 72 MHz to 37 GHz in Fig.~\ref{fig:SED}. We have chosen data points as close as possible to the time of the first MWA observation epoch (i.e., 2018 June 15). 
Due to the lack of low-frequency data in previous studies, the spectrum below 685 MHz has not been well constrained. The spectrum components below and above 685 MHz have distinctly different origins, so we fit the whole spectrum with two radiation components.
At $\nu>685$ MHz, the core dominates the emission, showing an inverted spectrum (or peaked spectrum);
at $\nu<685$ MHz, the flux density of the core is substantially reduced due to increasing synchrotron self-absorption (SSA) toward low frequencies, and the extended jets and lobes become gradually dominant.

We first fit the NE and SW spectra with power-law functions (i.e., $S \propto \nu^{\alpha}$) based on the VLA data from the literature \citep{2005A&A...435..497B} and obtained their spectral indices as $\alpha_{\rm NE} = -0.79$ and $\alpha_{\rm SW} = -0.72$, respectively. These are typical values for optically thin synchrotron emission. 
We then extrapolated the flux densities of NE and SW to the MWA band (central frequencies of 88, 118, 185, and 216 MHz). Next, the flux densities of the core were fitted with a self-absorbed synchrotron spectrum model \citep{1970ranp.book.....P,1999A&A...349...45T},  by fixing the optically-thick spectral index $\alpha_{\rm thick}$ to 2.5 (Appendix~\ref{app:spectrum}).
The fit yields a low-frequency power-law spectrum with the amplitude $A = 27^{+12}_{-9} $ mJy, the low-frequency spectral index $\alpha_{\rm low} = -0.97^{+0.21}_{-0.21}$, the baseline flux density $B = 41^{+8}_{-12} $ mJy, and a higher-frequency optically thick spectrum with the amplitude $F_m = 146^{+11}_{-9}$ mJy, the turnover frequency of $\nu_m = 11.24^{+2.01}_{-1.25}$ GHz, the optically-thin spectral index $\alpha = -0.20^{+0.10}_{-0.12}$. 
However, our spectrum shown in Fig.~\ref{fig:SED}, constructed with data points measured when the source was in a low-activity state (indicated by the red-colored arrow in Figure \ref{fig:lightcurve}), is different from that presented in \citet{1999ApJ...514L..17F} which was in the flaring state.  During the 1998 flare, the core had an inverted spectrum between 1.4 and 666 GHz peaking around 43 GHz \citep{1999ApJ...514L..17F}, a factor of 3 higher than our fit. In the quiescent state in 2018 depicted by Figure \ref{fig:SED}, the emission at GHz frequencies is still dominated by the core C, but the turnover frequency has shifted to $11.24^{+2.01}_{-1.25}$ GHz. 

We then subtracted the extrapolated flux densities of C, NE, and SW from the observed MWA flux densities, and the remaining  flux density is mainly from the extended components N+S. Finally, we fitted the N+S flux densities with a power-law function and obtained a spectral index of $\alpha = -1.09 \pm 0.12$, which is much steeper than that of the inner lobes NE and SW, but consistent with the spectral indices of radio relics \citep{2015A&A...583A..89S,2021PASA...38....8Q}.

\subsection{Temporal evolution of the VLBI components}\label{sec:temporalevo}

Figure \ref{fig:VLBI-2} panels \textit{c} and \textit{d} show the correlation between the core size and the flux density: the core size gradually increased when the flux density was in the rising phase before the flare peak; after the flare peak, the core size gradually decreased. The value of Spearman's rank correlation coefficient \citep{lehman2005jmp,myers2013research,dodge2008concise} between the core size and flux density is 0.43 with $p$-value of 0.14. Their positive correlation needs to be verified by more data. If this correlation is further confirmed, this phenomenon may be naturally explained by the superimposition of a flaring and fast-moving jet component on a quiescent and stationary core. The resolution of the VLBI images in this paper is not yet sufficient to distinguish between these two components in the initial flaring phase. Only when the jet knot produced after the flare moves to a certain distance is it sufficient to separate the jet from the core clearly. 

As the flaring component (a propagating shock) passed through the core, it led to a gradual increase in the flux density of the observed core. As the duration of the radio flare (lasting several years) was much longer than the cooling time of the synchrotron radiation of relativistic electrons ($\sim 130$ days at 15 GHz), it required a continuous injection of fresh relativistic electrons into the core to ensure that the core flux density continued to increase for about two years. The flaring component continued to move outwards and gradually separated from the stationary core. 
For a period after the flare (about 1 year), the VLBI image resolution was not sufficient to distinguish between the core and the flaring jet component, but the core size was observed to increase as the jet moved outward (Figure \ref{fig:VLBI-2}-c).  
With the cessation of the intermittent energy injection, 
the core returns to the quiescent state until the next flaring component passes by.  
At the same time, the size and brightness of the core gradually became smaller as the flaring jet component continued to move outward and separated from the stationary core. On the other hand, the flaring component became optically thin and its surface brightness declined over time. 
An alternative model of an inflating balloon has been proposed to explain the spectrum and structure evolution of \obj\ by 
\citet{1999ApJ...514L..17F}. In their model, the initial phase of the flare is interpreted as the relativistic jet interacting with the torus, and the jet--ISM collision causes frustration with the jet's advancing motion. Their model is compatible with the one we propose here to explain the evolution of major flares. Moreover, their jet--ISM explanation is also consistent with the jet-wind collision model we propose in Section \ref{sec:jet-wind} to interpret the secondary 2010 flare. 
High-cadence high-resolution VLBI monitoring of the flaring jet helps to refine this physical picture.

The flux density variability of the jet in Figure~\ref{fig:VLBI-2} panel c shows a delay relative to the core although this characteristic is not very pronounced due to the sparse distribution of the data points. On the other hand, the temporal evolution of the jet component size (see Table \ref{table2}) displays an inverse correlation with the core size, as is most evident after 2009 (corresponding to J2). 
We suggest that after separation from the core region, the moving discrete jet component underwent adiabatic expansion, leading to an increase in size, as well as a decrease in surface brightness. The jet disappeared until the brightness of the shock fell below the detection threshold.

To summarize the properties of the jet of \obj\ described above, we find that the radio properties of \obj\ in the quiescent state make it look more like a Seyfert 1 galaxy.
However, its strong variability, apparent superluminal jet motion, and large extended structure are very distinct from normal RQ quasars, but instead, resemble RL quasars. 
Especially, its radio properties during the major flares, such as superluminal jet motion and high brightness temperature, are consistent with those of a blazar \citep{1999ApJ...514L..17F}. This hybrid feature, being a radio-quiet quasar at the quiescent state but behaving like a blazar at the flaring state, has been noted in previous studies \citep[e.g.][]{1999ApJ...514L..17F} and 
might be a common feature of radio-intermediate AGN (Section \ref{sec:RIquasar}).

\subsection{Periodic flares and jet ejection}\label{sec:variability}

\begin{figure*}
    \centering
    \includegraphics[width=0.45\textwidth]{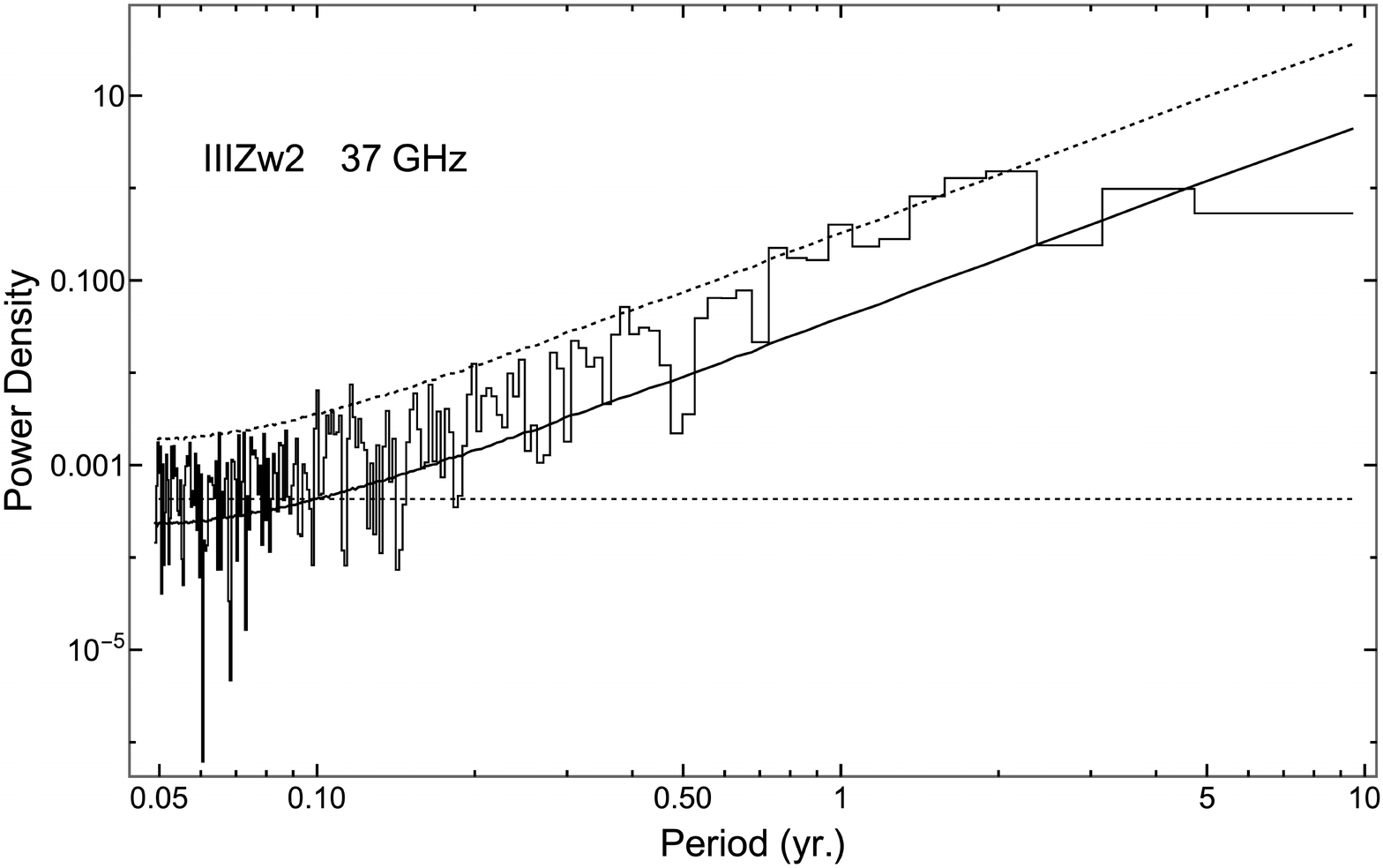}
    \includegraphics[width=0.45\textwidth]{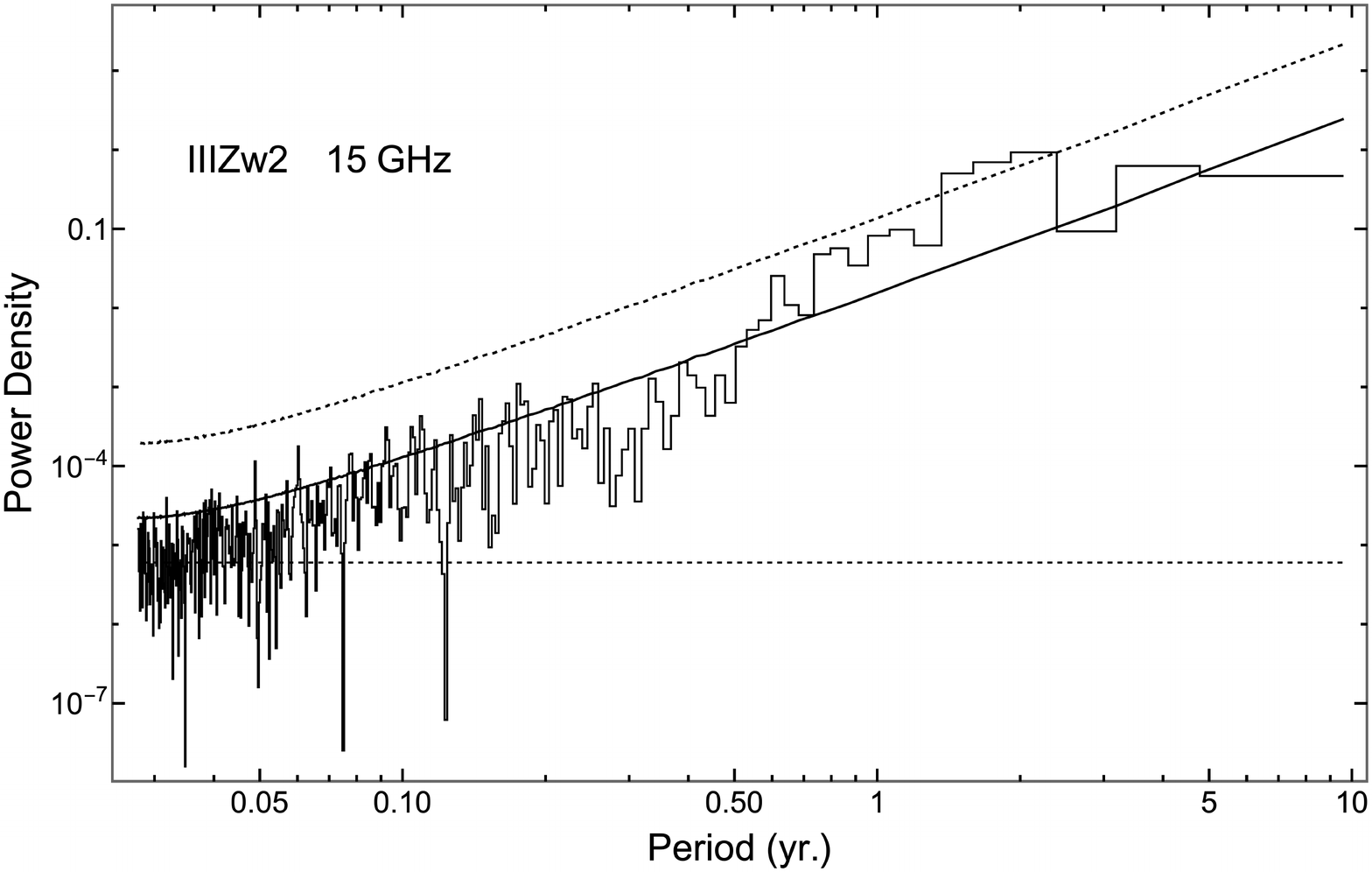} \\   
    \includegraphics[width=0.45\textwidth]{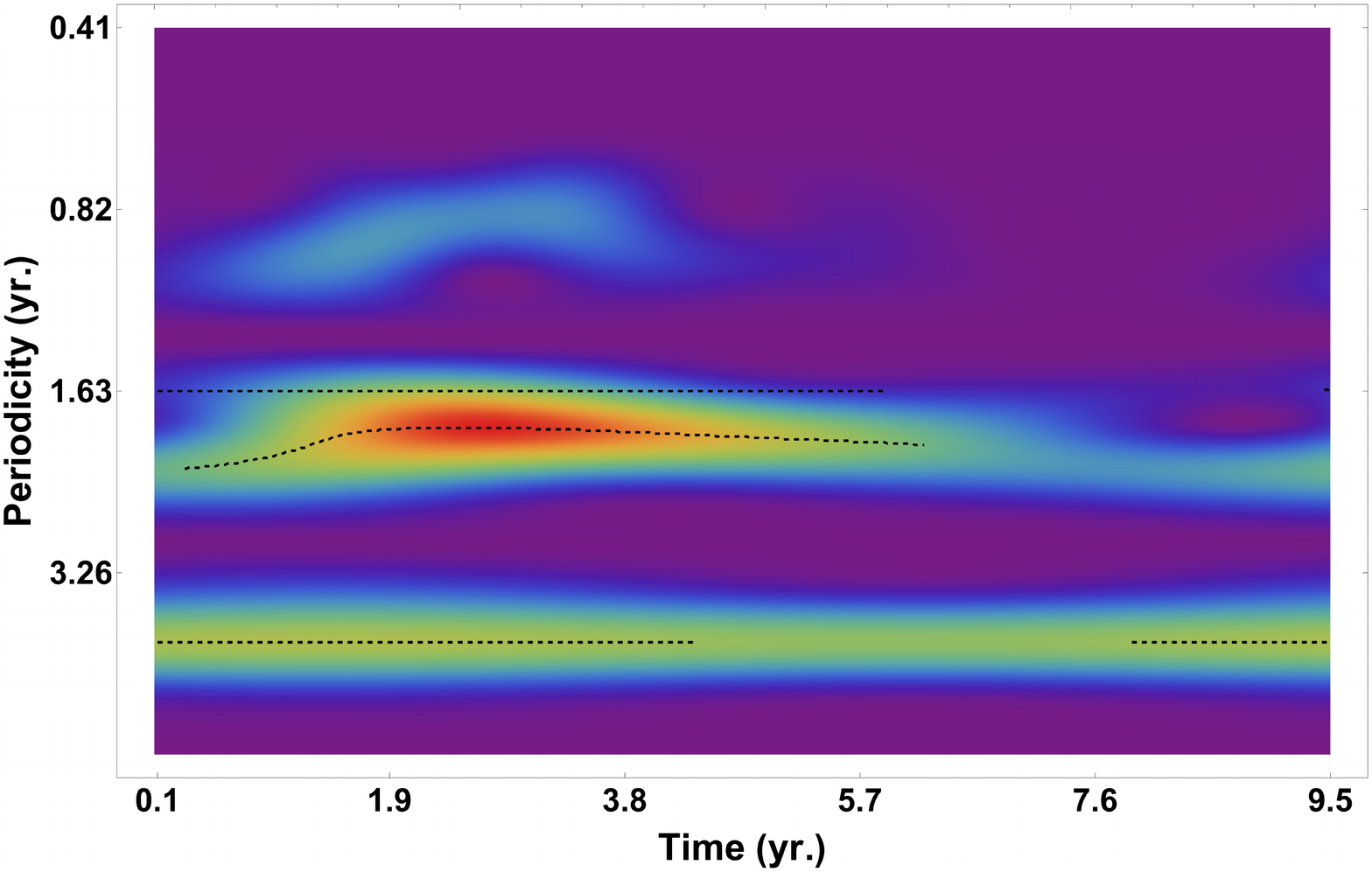}    
    \includegraphics[width=0.45\textwidth]{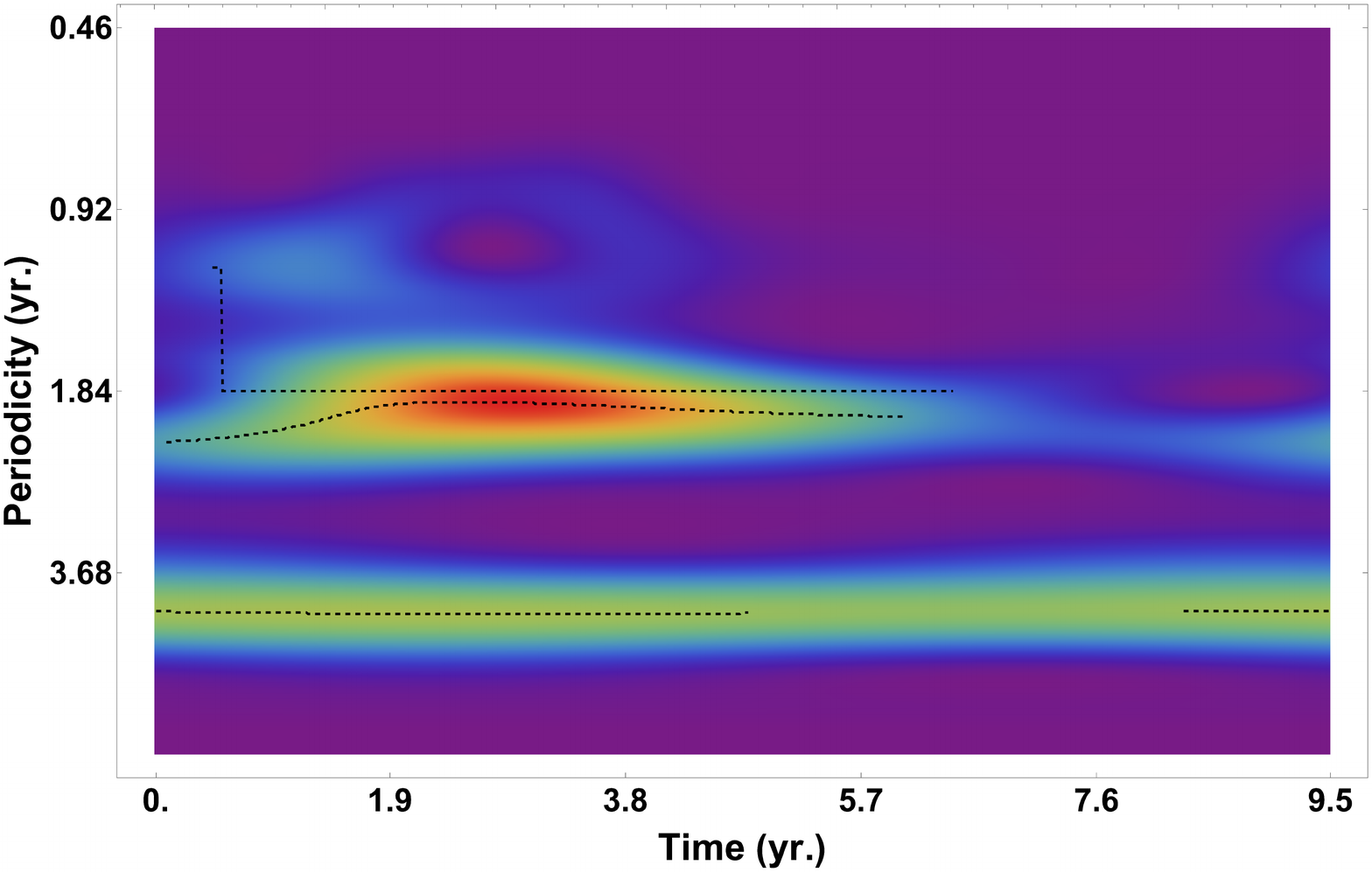} 
    \caption{Periodogram of \obj. The top panels show the periodograms of 37 GHz (\textit{left}) and 15 GHz light curves (\textit{right}). In the periodogram plots, the horizontal dashed line is the level of white noise (constant). The curved line represents the mean periodogram (closest to the underlying power spectral density) from the Monte-Carlo simulations using the best-fit model parameters; this additionally accounts for model uncertainties. The dashed curve is the 95 \% confidence level (encompasses 95\% of the periodogram ordinates; any outliers are statistically significant quasi-periodic oscillations). A broad peak stands out in both plots, corresponding to a period of $\sim 2$ yr. The bottom panels show the results from wavelet analysis of the light curves from 2011.6 to 2021: {\it left} -- 37 GHz, {\it right} -- 15 GHz. A persistent component corresponding to a characteristic timescale of 1.97--2.02 yr is clearly seen throughout the time span. A secondary feature appears at a timescale of 3.97--4.26 yr.  }
    \label{fig:periodogram}
\end{figure*}

Figure~\ref{fig:lightcurve} shows the 37-GHz light curve with two particularly major flares in the time period 2003--2020, peaking in late 2004 and late 2009, respectively. The 2004 flare has been reported by \citet{2010NewA...15..254L}. The 2009 flare is the highest in magnitude in $\approx$40 years of radio monitoring, with a maximum flux density of 3.12 Jy, approximately 20 times the baseline in the quiescent state. Such a large radio variability is rare even in blazars' light curves \citep{2011ApJS..194...29R}.
After the 2009 flare, the source became less active and the flare peak gradually decreased, although there were a few lower-amplitude flares in 2013, 2015, and 2017. Since 2018.4, the source has entered a quiescent state. 
Similar to the 37-GHz light curve, the 15-GHz light curve also displays a number of flares with the largest one in late 2009 -- early 2010. The maximum flux density increased by a factor of 10.5 compared to the baseline level. After 2016, the source entered a low-level state.

\cite{2003PASA...20..126B} found that \obj\ has major  radio flares about every five years. \cite{2010NewA...15..254L} found the same flare period 
based on the historical light curves of \obj\ at 22 and 37 GHz obtained from the Mets\"{a}hovi Radio Observation database \citep{2005A&A...440..409T} covering 18 years from 1986 to 2004. 
We continue to analyze the periodicity of the 37-GHz light curve from 2011.40 and the 15-GHz light curve after 2008 using the Lomb-Scargle periodogram, yielding  a period of $P\sim2.1$ yr (Appendix \ref{app:singledish}). 
This periodic signal is not sharply distributed in the Lomb-Scarge periodograms, therefore it can only be called a quasi-periodic signal. In the wavelet periodograms, the most prominent feature is around 1.97--2.02 yr (Figure \ref{fig:periodogram}), and a secondary weaker feature is around 3.97-- 4.26 yr. The period of $P \sim 4$ yr has been steadily maintained for $\sim$35 years (\citealt{2005A&A...435..497B,2010NewA...15..254L} and the present paper) with 6--7 complete cycles, suggesting that the periodicity should not be a fake signal caused by random red noise, but is related to some intrinsic dynamical process. The 2-yr periodicity has also appeared in previous periodicity analysis \citep{2010NewA...15..254L}, but the magnitude is relatively low. 

Quasi-periodic variations in AGN light curves are a common observational phenomenon and are often interpreted as being related to regular perturbations at the jet origin, such as the precession of the jet nozzle 
\citep{2008Natur.452..851V} and rotation of a hotspot along a helical path \citep{1992A&A...255...59C,2016MNRAS.463.1812M}, or magnetohydrodynamic or hydrodynamic instabilities arising from the starting section of the jet \citep{2003ApJ...597..798H,2022Natur.609..265J} or the disk \citep{2013MNRAS.434.3487A,2014MNRAS.443...58W} and propagating as helical-mode waves.

In many cases, warping of the accretion disk can occur. In the case of jet precession, if the spin axis of the primary black hole is not aligned with the orbital plane of the binary, the differential precession with the change in radius can cause the warping of the outer disk \citep{1975ApJ...195L..65B}. Alternatively, the self-irradiation of a luminous accretion disk can also lead to the disk warping out of the orbital plane \citep{1996MNRAS.281..357P}. 
The dynamic process associated with the warped disk generally occurs in the outer disk region, while the jet is launched from the inner region of the disk \citep{1982MNRAS.199..883B} or the vicinity of the black hole \citep{1977MNRAS.179..433B}. There is a big gap between the two physical processes of jet launching and disk warping. The coupling of the jet and disk would require that the instabilities originating from the edge or the outer region of the disk can be transmitted to the inner region, whereas this inward propagation is difficult to achieve through viscous processes because the required time scale is too long. Instead, the yr-timescale periodic variability of \obj\ could result from global acoustic or $p$-mode oscillations in a thick disk \citep{2003MNRAS.344L..37R, 2003MNRAS.344..978R}. The inferred variability period is on the order of several years for a black hole mass of $10^8 M_\odot$ \citep{2013MNRAS.434.3487A}.
The trigger of the $p$-mode oscillations can be a locally periodic agent at the edge of the disk due to either gravitational or radiation perturbations. 
The inward propagation of global oscillations to the inner region induces quasi-periodic fluctuations in the accretion flow, which in turn trigger quasi-periodic injection of plasma into the jet, leading to the observed harmonic periodic radio variability and the corresponding periodic jet ejection.

In addition to the primary 4-yr period, there is a secondary peak at $P \approx 2$~yr in the periodograms, consistent with four lower-amplitude flares in 2013, 2015, 2017 and 2019 (Figure~ \ref{fig:lightcurve}). The 2-yr periodic component could be a harmonic of the 4-yr periodicity, with each being dominant in different activity states. 
Only the fundamental frequency of the oscillations (i.e. $f = 0.25 \, {\rm yr}^{-1}$), can produce the largest flare as well as the long-lasting jet knot which can be observed in VLBI images. Lower-amplitude flares associated with harmonic components may also generate jet knots, but they are too weak to be detected. 
The presence of multiple harmonic periodicities on years timescale can be explained by hydrodynamic instability, such as the global \textit{p}-mode oscillations of the accretion disk. 
Moreover, the quasi-periodic signals induced by the accretion disk instability have no fixed frequency and often vary within a range, which is consistent with the observations.

\subsection{Correlation between gamma-ray and radio flares} 

There are correlations between the prominent variabilities of \obj\ in multiple bands (see Introduction).
\cite{2016ApJS..226...17L} identified two $\gamma$-ray flares that peaked in 2009 November and 2010 May, respectively.
The properties of \obj\ $\gamma$-ray flares are analogous to those of blazar flares \citep{2016ApJS..226...17L}, implying that the central engines between \obj\ and blazars are not essentially different.
The radio light curves (Figure \ref{fig:lightcurve}) also have substructures during the period 2009--2011: in addition to the strongest flare peaking in late 2009 (2009 flare, hereafter), there is also a lower-amplitude flare on the shoulder of the declining phase that peaks mid-2010 (2010 flare, hereafter).

The 2009--2011 lightcurves could be better described with two flare components than with just one (Section \ref{app:2009flare}).
The observed light curve matches well with a simulated light curve consisting of two ``exponential rise + exponential decay" flare components as an approximate description of the composite behavior of the flares (Figure \ref{fig:2009flare}). 
At both frequencies, the primary 2009 flare is brighter than the secondary 2010 flare. The 2009 flare peaks at epoch 2009.96 at 37 GHz, leading the peak epoch 2010.09 at 15 GHz by $\sim$47 days.  The 2010 flare peaks at epoch 2010.56 at 37 GHz, and at epoch 2010.58 at 15 GHz, with no significant time delay. 
The fact that the 37-GHz radio flare leads the 15-GHz flare can be naturally explained by the frequency-dependent opacity \citep{2008A&A...485...51H}.
The time span between the 2009 and 2010 flares of 0.5--0.6 yr is much longer than the lifetime of synchrotron electrons (i.e., the cooling time of 85 days and 135 days at 37 and 15 GHz), but much shorter than the time separation between two major flares ($\sim4$ yr). Therefore, the 2009 and 2010 flares must be associated with the same episodic nuclear activity but are unlikely to be the same emission components. 

The 2009 flare is reminiscent of the typical core flares occurring in blazars. In $\gamma$-ray AGN, both the $\gamma$-ray flare and the associated delayed millimeter-wavelength flare are created in the standing shock in the innermost jet (the shock-in-jet model \citealt{2008Natur.452..966M}), which is optically thick and shows time delays in the light curves at different radio frequencies.
The 2009 radio flare lagged the $\gamma$-ray peak (2009.84) by only $\sim$40 days \citep{2016ApJS..226...17L}, suggesting that the $\gamma$-ray emitting zone is spatially connected to the radio-emitting zone and the $\gamma$-ray emitting site is closer to the central engine than the radio flare site.   
In contrast, the concurrent 2010 flare at 37 and 15 GHz frequencies suggests that it should arise from an optically-thin jet component, probably associated with a compressed shock in the jet propagating downstream. 
In addition, the time-integrated flux density is also higher at 37 GHz than at 15 GHz, implying that the energy output of the radio source is concentrated at higher frequencies during the most active phase of the flare. Whereas, the peak time and maximum amplitude of the 2010 flare are similar at both 37 GHz and 15 GHz, suggesting that the energy dissipation of the 2010 flare is weakly dependent on frequency, reinforcing the intrinsic difference between the 2009 and 2010 flares. 

\subsection{Jet-wind collision and the associated 2010 flare}\label{sec:jet-wind}

The pieces of observational evidence presented in previous sections, including the temporal evolution of the flux density and jet component size, optically-thin radio spectrum, and correlated radio/$\gamma$-ray flares, lead us to speculate that the 2010 flare could be a compressed shock caused by the downstream jet-ISM collision. 
Additional evidence for jet-ISM collision comes from the VLBA MOJAVE polarimetric observations: the fractional linear polarization of \obj\ increased from 0.2\% on 3 June 2009 (prior to the 2019 flare) to 0.7\% on 12 July 2010 (corresponding to the peak time of the 2010 flare), and the polarization angle changed from $6^\circ$ to $24^\circ$.
The enhanced linear polarization and change of polarization angle in the jet offer direct evidence of a compressed shock created in the jet-ISM collision, as has been found in other radio-loud quasars \citep{2020NatCo..11..143A}.

In the warped disk model discussed in Section \ref{sec:variability}, the wind or outflow driven by the radiation pressure of the disk is released from the outer region of the disk to form a cylindrical or conical structure in the broad line region (BLR). The jet flushes a tunnel in the axial direction of the wind, in which the jet moves outwards.  The axis of the disk wind is bound to the axis of the outer disk, and the oscillations of the outer disk would lead to the wind wall oscillating within a certain angle as well. Thus, the jet axis is not always parallel to the wind axis; at a certain distance, the jet may hit the inner boundary of the wind wall, producing an oblique shock. On the jet-wind interaction interface, the oblique shock is deflected, after which the shock (jet knot) follows a (new) ballistic trajectory in the direction of the deflection. The loss of the jet kinetic energy during this collision leads to the dissipation of radiation, producing the observed prominent $\gamma$-ray and radio flares (e.g., the 2010 flare). On the other hand, the oscillations of the wind boundary would cause the working surface of the jet-wind collision to vary with time, which seems to coincide with the observed change in the position angle leading to a change in the position angle of the (deflected and redirected) VLBI jet, as seen in Figure \ref{fig:VLBI-2}.

Recent studies \citep{2016A&A...585A..33B,2021A&A...647A..67B} have shown that the powerful nuclei of high-excitation galaxies produce disk-launched winds/outflows which could form the slower jet sheaths, in addition to highly relativistic jet spines. The slower sheath may contribute to the collimation of the jet. This is compatible with the model we propose: in our model, it is the accretion disk wind rather than the outer-layer jet that surrounds the jet spine. Collision may occur between the fast-moving jet spine and the inner boundary of the disk wind when the axis of the winds is not aligned with the jet axis.

\textit{Where did the flares occur?} Extrapolating the trajectory of the jet component J2 back to the peak times of the 2009 and 2010 radio flares, we obtain a distance of J2 of 0.096 mas (a projected distance of 0.16 pc) and 0.205 mas (0.34 pc) from the black hole, respectively.  Neither of these size scales is resolvable by the current 15 GHz VLBI observations, and even the previous 43 GHz VLBI observations (resolution of 0.1 mas, \citealt{2005A&A...435..497B}) can only barely resolve this size. Therefore, all relevant jet--wind collision and flaring activities are hidden in the 15-GHz VLBI core. 
Using the same method, we extrapolate to estimate the distance of the 2009 $\gamma$-ray flare site from the central black hole to be 0.068 pc, a distance that can be considered as the upper limit of the jet collimation zone. That is to say, the jet collimation must have been complete at this distance ($7.6\times10^3 R_g$ in projection). This distance is comparable with that of the jet collimation zone derived from other AGN, such as M87 \citep{2012ApJ...745L..28A,2013ApJ...775...70H} and other nearby radio galaxies \citep{2021A&A...647A..67B}.  
The 2010 $\gamma$-ray flare site, $\lesssim0.27$~pc from the central engine, favors that the jet-wind collision occurs farther downstream in the jet.

\subsection{Comparision with other similar sources} \label{sec:RIquasar}

In this section, we compare the radiation and physical properties of \obj\ with Mrk~231 and explore the generalized properties of RI AGN.
Mrk~231 is one of the closest known radio-quiet quasars with an extremely strong infrared luminosity and rich multi-phase multi-scale outflows 
(\citealt{2021MNRAS.504.3823W} and references therein). \obj\ shares similar observational features with Mrk~231: ongoing galaxy mergers, high luminosity, misalignment between the pc-scale and kpc-scale jets, prominent flares, and the associated intermittent jet ejection. 
In addition, both sources behave like radio-quiet AGN in the quiescent state, while like a blazar in the flaring state \citep{2023MNRAS.518...39W}. Their flare properties and jet kinematics can be explained by the classical synchrotron self-absorption radiation model. The jet knots follow ballistic trajectory on a few parsec scales, but the position angle varies from one knot to another. These variability properties and jet structure change are reflective of interactions between the jet and the interstellar medium in the broad line region, or the jet being reflected by a rotating torus \citep{2021MNRAS.504.3823W}. 
\obj\ differs from Mrk~231 in that the \obj\ jet extends to a larger distance, while the Mrk~231 jet is confined within the host galaxy. The ability to develop large-scale jets depends not only on the initial kinetic properties of the jet, but also on the external environment (whether or not it chokes the jet), and the ability of the central engine to remain active for a long period has a profound effect on the jet growth \citep{2012ApJ...760...77A}.

\section{Summary}

In this paper, we analyze in detail the radio structure and variability properties of \obj, a radio intermediate AGN. The main results are summarized below:
\begin{itemize}
    \item The overall jet structure from pc to kpc scales shows an S-shaped morphology, probably related to the jet re-orientation due to galaxy interaction. The low-frequency ASKAP and MWA images confirm the presence of extended emission $27^{\prime\prime}$ to the north and $26^{\prime\prime}$ to the south from the core. The ultra-steep spectra of these extended features suggest that they are relics of past AGN activity.
    
    \item Two jet components J1 and J2 are detected in the VLBI images, with an apparent superluminal velocity of $1.35\,c$ and an average jet viewing angle of $\sim 20\degr$.

    \item The radio light curves show quasi-periodic flares: before 2008, a $\sim$4-yr cycle dominates; after 2008, when the source is in a low-activity state, a high-frequency harmonic component of $\sim$2-yr period becomes dominant. The variability characteristics (quasi-periodicity, two periodic signals, and the presence of harmonic relation between them) can be explained by the global acoustic oscillations of the accretion disk. The perturbations occurring in the outer region of the disk propagate inward, leading to modulated changes in the accretion rate and, consequently, to the generation of periodic radio flares and jet ejection. 
    
    \item  Only major flares associated with the fundamental frequency oscillation can produce observable jet components. The two strongest flares occurring in late 2004 and late 2009 coincide with the creation of the jet knots J1 and J2 observed in the VLBI images, respectively. 
    
    \item The radio flare from late 2009 to mid-2010 can be decomposed into two sub-flares, corresponding to two $\gamma$-ray flares, respectively. The 2009 $\gamma$-ray flare led the radio flare and the high-frequency radio flare led the low-frequency flare, suggesting that it originated from an optically thick component, probably in the jet collimation region. While the 2010  radio flare occurred simultaneously at 37 and 15 GHz, suggesting that they occurred in an optically thin jet zone.

    \item  The wind or outflow arising from the outer accretion disk forms a cylinder or cone in the nuclear region. As the axis of the warped disk is misaligned with the jet. At a certain distance, the jet flow hits the wind wall, creating an oblique shock that deflects the jet; at the same time, the jet-wind collision leads to the production of $\gamma$-ray and radio flares (e.g., those observed in 2010). 
    
\end{itemize}

\obj\ has hybrid nature of RQ AGN and blazar: in the quiescent state, it is a typical RQ AGN, while in the flaring state it behaves as a blazar.  During the intermittent flares, the produced jet knots interact with the accretion disk wind in the broad line region, producing $\gamma$-ray and radio flares.
The characteristics observed from \obj\ may be common to the RI AGN population, in which both the jets and winds coexist and play important roles at different spatial scales and timescales. 
Detailed studies of typical individual RI AGN will improve our understanding of the RQ/RL AGN dichotomy and the structure and dynamics of the AGN nuclear region.

\section*{Data Availability}

The datasets underlying this article were derived from the public domain in NRAO archive (project codes: BU013, BA080;
\url{https://science.nrao.edu/observing/data-archive}), Astrogeo archive (project codes: BB023, RDV13, BG219D, UF001B, UG002U;
\url{http://astrogeo.org/}), the MOJAVE data can be found from the MOJAVE website (\url{https://www.cv.nrao.edu/MOJAVE/sourcepages/0007+106.shtml}), MWA archive (\url{https://asvo.mwatelescope.org}) and CSIRO ASKAP Science Data Archive (CASDA, \url{https://research.csiro.au/casda/}). The MWA GLEAM-X data is not publicly released and the calibrated visibility data can be shared on reasonable request to the corresponding authors. Data from the Owens Valley Radio Observatory and the Mets\"{a}hovi Radio Observatory can be requested from the respective data maintainers.


\begin{acknowledgments}
This work was supported by resources provided by the China SKA Regional Centre prototype \citep{2019NatAs...3.1030A,2022SCPMA..6529501A} funded by the Ministry of Science and Technology of China (MOST; 2018YFA0404603). 
This research has been supported by the National SKA Program of China (2022SKA0120102).
S.G. is supported by the CAS Youth Innovation Promotion Association (2021258).
The authors acknowledge the use of Astrogeo Center database maintained by L. Petrov. The National Radio Astronomy Observatory are facilities of the National Science Foundation operated under cooperative agreement by Associated Universities, Inc.
This publication makes use of data obtained at the Mets\"{a}hovi Radio Observatory, operated by the Aalto University in Finland.
This research has made use of data from the MOJAVE database that is maintained by the MOJAVE team \citep{2018ApJS..234...12L}.
This work makes use of the Murchison Radioastronomy Observatory, operated by CSIRO. We acknowledge the Wajarri Yamatji people as the traditional owners of the Observatory site. Support for the operation of the MWA is provided by the Australian Government (NCRIS) under a contract to Curtin University, administered by Astronomy Australia Limited. 
We acknowledge Paul Hancock, Gemma Anderson, John Morgan, and Stefan Duchesne for their contributions in GLEAM-X pipeline which bring great convenience to us.
This research has made use of data from the OVRO 40-m monitoring program which was supported in part by NASA grants NNX08AW31G, NNX11A043G and NNX14AQ89G, and NSF grants AST-0808050 and AST-1109911, and private funding from Caltech and the MPIfR.
This research has made use of the \citet{https://doi.org/10.26132/ned1}
which is funded by the National Aeronautics and Space Administration and operated by the California Institute of Technology.  
\end{acknowledgments}

%

\vspace{5mm}
\facilities{ASKAP, GMRT, HST, MWA, VLA, VLBA}


\software{AIPS \citep{2003ASSL..285..109G}, astropy \citep{2013A&A...558A..33A,2018AJ....156..123A}, Difmap \citep{1994BAAS...26..987S}
          }



\clearpage

\appendix
\restartappendixnumbering

\section{Low-frequency observations of III Zw 2} \label{app:MWAdata}

We study the low-frequency radio structure of \obj\ based on the recent observations from the Murchison Widefield
Array (MWA, \citealt{2013PASA...30....7T}), the Giant Metrewave Radio Telescope (GMRT, \citealt{1991CSci...60...95S}) and the Australian SKA Pathfinder (ASKAP, \citealt{2007PASA...24..174J,2014PASA...31...41H}). 

The MWA is the precursor of the Square Kilometre Array (SKA) low-frequency telescope and is located in Western Australia. 
We have downloaded the latest unpublished data of \obj\ observed by the MWA Phase II \citep{2018PASA...35...33W} on 2018~June~15.  The extended array of the MWA phase II  comprises 128 tiles (each consisting of 16 crossed-pair dipole antenna), distributed over a maximum baseline of $\sim$6 km, twice longer than that of Phase I. These data are included in the GaLactic and Extragalactic All-sky MWA survey -- eXtended (GLEAM-X, \citealt{2022PASA...39...35H}). The GLEAM-X survey covers the entire sky south of declination $+30^\circ$ and at the same frequency range of 72--231 MHz as GLEAM \citep{2017MNRAS.464.1146H}. GLEAM-X achieves a typical sensitivity of $1\sigma \approx 1.3$ \mJyb\ in the frequency range of 170--231 MHz, which is eight times more sensitive than GLEAM.  We downloaded 40~observation snapshots containing \obj\ from the MWA archive \footnote{\url{https://asvo.mwatelescope.org}}, and processed the data at the China SKA Regional Centre \citep{2019NatAs...3.1030A,2022SCPMA..6529501A} using the pipeline developed by the GLEAM-X team\footnote{\url{https://github.com/tjgalvin/GLEAM-X-pipeline} maintained by Tim Galvin}. For each snapshot, we first flagged the bad tiles or receivers and obtained reliable calibration solutions. Next, we flagged potential radio frequency interferences. The calibration solutions obtained from the calibrators were then applied to the data.  A Briggs robust parameter was set to $+0.5$ in imaging to maximize sensitivity. We made a deep \textsc{clean}ed image, followed by source-finding, ionospheric de-warping, and flux density scaling to GLEAM in the \textsc{clean} images. Then the  point spread function (PSF) placeholder of each snapshot image was corrected. In the last step, we combined the astrometrically- and primary-beam-corrected snapshot images into a high signal-to-noise image with reduced noise, allowing for detecting fainter sources and diffuse structures that are not visible in individual snapshot images.

\begin{figure*}
    \centering
    \includegraphics[width=1.0\textwidth]{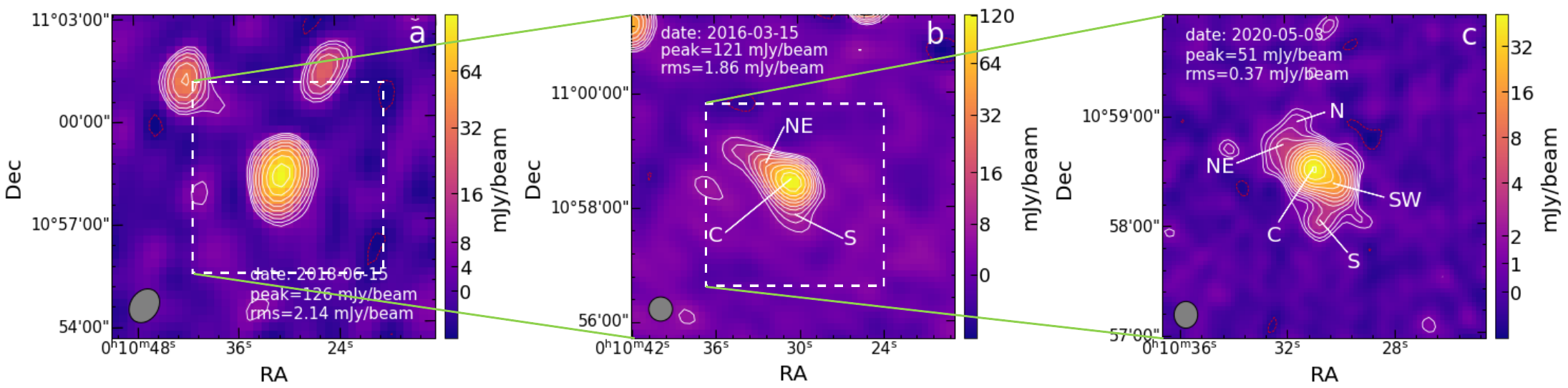}
    \caption{Radio images of \obj. 
    (a): the 215.5-MHz image observed from GaLactic and Extragalactic All-sky Murchison Widefield Array survey eXtended (GLEAM-X) survey on 2018 June 15. The peak contour flux is 126 \mJyb\  and the contour levels are 6.42 $\times$ ($-1$, 1, 1.4, 2, 2.8, 4, 5.6, 8, 11.2, 16, 23, 32) \mJyb. The \textit{rms} noise is 2.14 \mJyb. The beam size is $63\farcs1 \times 49\farcs1$ with the major axis along a position angle of $146\fdg8$. 
    (b): the 150 MHz image acquired the TIFR GMRT Sky Survey Alternative Data Release \citep[TGSSADR, ][]{2017A&A...598A..78I} observed on 2016 March 15. The integration time of each pointing is $\sim$15 min with the beam size of $25\arcsec\times25\arcsec$. 
    (c): the 888 MHz image obtained from the Rapid ASKAP Continuum Survey \citep{2020PASA...37...48M}. The observation was made in 2020 May 03. The integration time on the source is $\sim$15 min. The peak intensity is 51 \mJyb. The lowest contour level is 1 \mJyb, and the contours increases in the step of $\sqrt{2}$. The \textit{rms} noise in the image is 0.33 \mJyb. The overall size is 64\farcs1 ($\sim$ 106 kpc). The color scale shows the intensity in the logarithmic scale. }
    \label{fig:mor}
\end{figure*}

Figure~\ref{fig:mor}-a shows the 216-MHz MWA image of \obj, displaying a compact component. 
The resolution of the image is $63\farcs1\times49\farcs1$, and the root-mean-square (\textit{rms}) noise is 2.1 \mJyb\ (a factor of $\sim4.8$ lower than that of the GLEAM image). The \textit{rms} noise of the \obj\ image is 1.7 times higher than the mean value of the GLEAM-X images, due to the lower elevation angle of the MWA when observing \obj\ at $\delta=+11^\circ$.  \obj\ is marginally resolved and shows an elongation along the north-south direction. The source is unresolved at other lower frequencies.
The GLEAM-X data points themselves can be fitted with a power law with a steep spectral index $\alpha = -0.96 \pm 0.09$. 

Figure \ref{fig:mor}-b shows the GMRT image at 150 MHz, showing a resolved structure along the NE-SW direction with a total extent of  $\sim 48\arcsec$.
The GMRT image of \obj\ is from the TIFR GMRT Sky Survey Alternative Data Release (TGSS-ADR1 \footnote{\url{https://vo.astron.nl}}) \citep{2017A&A...598A..78I}. TGSS-ADR1 covers 90 per cent of the total sky from $\delta = -53^{\circ}$ to $\delta = +90^{\circ}$ with a noise level of $\sim$5 \mJyb\ and a resolution of $25\arcsec\times25\arcsec$/cos(Dec$-19^{\circ}$) at 150 MH. 
The total flux density is $\sim$ 212 mJy, dominated by the central core.
Besides the core C, there are extensions toward the northeast (NE) and toward the south (S).

We obtained the ASKAP image of \obj\ from the Rapid ASKAP Continuum Survey (RACS), which is a shallow all-sky (covering the entire sky south of declination $\delta = +41^\circ$) pilot survey for future multi-year surveys with the full-scale ASKAP \citep{2020PASA...37...48M}. The observations covering the \obj\ position began on 2019 April 21 and lasted three weeks. The RACS has an instantaneous bandwidth of 288 MHz and is centered at 888 MHz. 
The angular resolution of the resulting image using the natural weighting is $14\farcs20 \times 12\farcs93$. The distance between NE and SW lobes is $\sim 33\arcsec$ and the image resolution enables to resolve the jet structure. The \textit{rms} noise in the image is 0.33 \mJyb, which is consistent with the mean \textit{rms} of the RACS images. 
The ASKAP image is shown in Fig.~\ref{fig:mor}-c, which reveals much richer details than the MWA image: the main jet body is elongated along the northeast--southwest (NE--SW) direction and consists of the core C and NE and SW lobes. The ASKAP image shows a very similar structure to the recently published uGMRT image in \citet{2020MNRAS.499.5826S}, although the two images are observed at different frequencies. As the observed frequency of the ASKAP image is higher than that of the uGMRT image, the emission from the outermost extended components beyond the NE and SW lobes (labeled as N and S in Fig.~\ref{fig:mor}-c) is fainter due to their steep spectrum nature.  
The separation between N and S lobes is $\sim55\arcsec$ ($\sim$98 kpc)\footnote{Adopting the following cosmological parameters: H$_0$ = 71 \kms\ Mpc$^{-1}$, $\Omega_\Lambda = 0.73$ and $\Omega_m = 0.27$, at $z = 0.0893$, 1 mas angular separation corresponds to 1.65 pc linear size in projection on the plane of the sky.},
larger than the size of the previously detected structure between NE and SW lobes in VLA images at GHz frequencies \citep{2005A&A...435..497B}. In the 685-MHz uGMRT map, the southern extension (labeled ML in \citealt{2021MNRAS.507..991S}) is much longer than our Figure~\ref{fig:mor}-c.
The total flux density estimated from the ASKAP images is 92.5 mJy, of which the main body (C+NE+SW) accounts for $\sim$86.3 mJy, while the remaining $\lesssim$6.7\% (6.2 mJy) of the flux density comes from the sum of the N and S lobes.

\begin{figure*}
\includegraphics[width=0.8\textwidth]{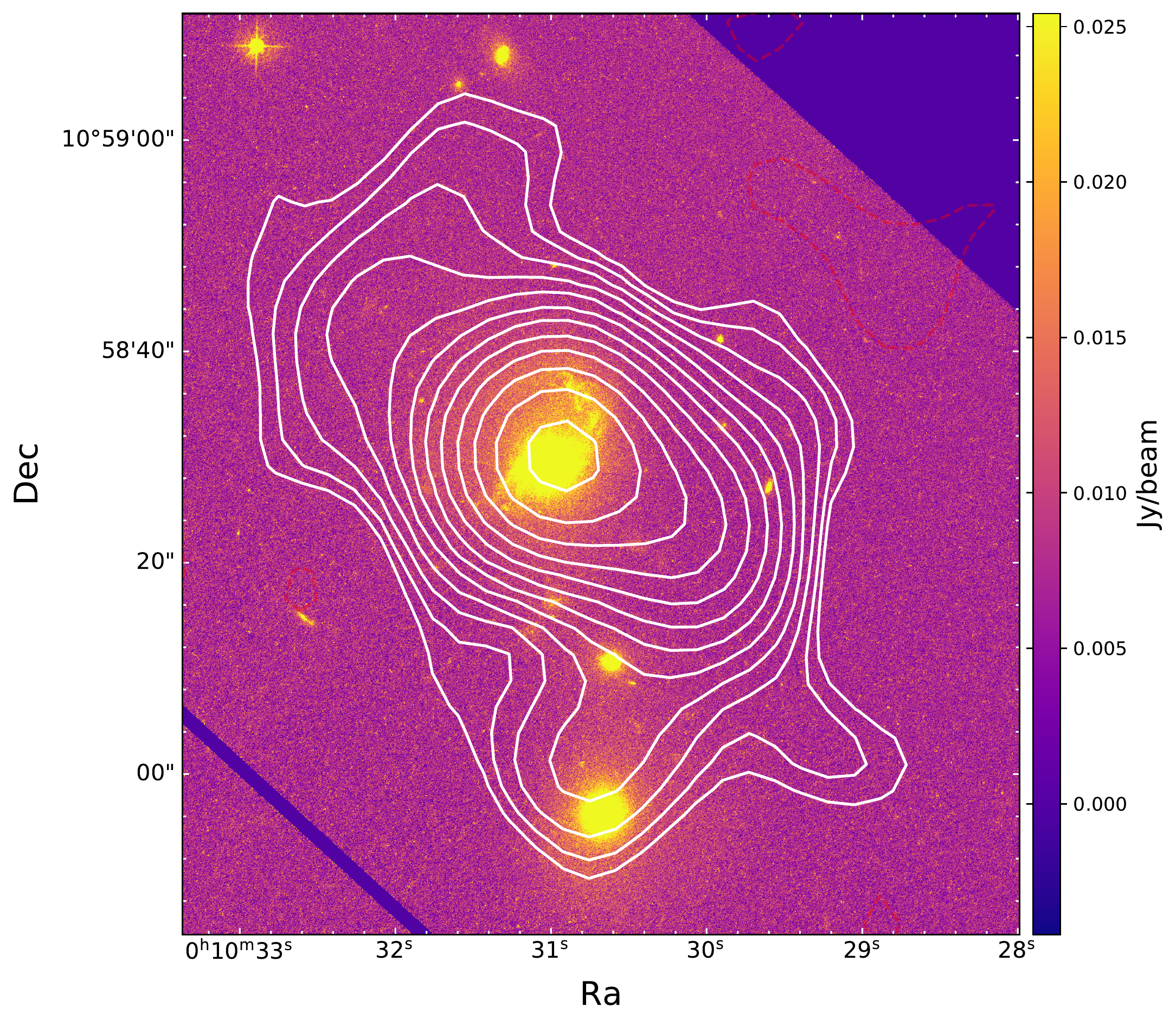}
\caption{Combined radio (ASKAP) and optical (HST) images. The ASKAP image is obtained from the Rapid ASKAP Continuum Survey (RACS) centered at 888 MHz. The HST image is acquired from Hubble Legacy Archive at the wavelength of 5446.8\AA.}
\label{fig:tidalarm}
\end{figure*}

The power-law portion at low frequencies originates from large-scale extended emission structures in the outermost N and S lobes which are evident from the MWA image of \obj\ and part of the structures are shown in the uGMRT \citep{2021MNRAS.507..991S} and ASKAP images. These steep-spectrum features correspond to aged structures, which become very faint above 1 GHz  \citep[e.g.][]{2017ApJ...836..174C}. 

The flux density from optically thin synchrotron emission is  \citep{1986rpa..book.....R,2013LNP...873.....G}
\begin{equation}\label{Fnuthin}
F_{\nu,{\rm thin}} = \frac{4 \pi R^3}{3 D^2_L} c_5 (p) K_0 B^{(p+1)/2} \left(\frac{\nu}{2 c_1} \right)^{-(p-1)/2},
\end{equation}
where $R$ is the size of the extended emission structure, $D_L$ is the luminosity distance, $K_0$ is a normalization factor, $B$ is the magnetic field strength, $\nu$ is the observing frequency, $p$ is the energy index, and  $c1$ and $c2$ are radiative constants \citep{1970ranp.book.....P}.   
Assuming a similar power law distribution of synchrotron emitting electron energies, $N (E)~dE = K_0 E^{-p}~dE$, and energy equipartition between the magnetic fields and particle kinetic energy densities, the normalization $K_0$ can be evaluated similarly to eqn. \ref{K0eqn}. The synchrotron frequency corresponds to the minimum injected Lorentz factor of emitting electrons and can be calculated as
\begin{equation}\label{eq:nu_m}
\nu_m = \frac{e B \gamma^2_m}{2 \pi m_e c}.
\end{equation}
Using eqn. \ref{eq:nu_m} and the normalization in eqn. \ref{Fnuthin}, and $\gamma_m = ((p-2)/(p-1)) (m_p/m_e) \epsilon_e$, in which $\epsilon_e$ is the fraction of the total energy density  in the emitting region in the magnetic fields.
The magnetic field strength in this region can be estimated as
\begin{align} \label{Blargescale}
B &= \left(\frac{6 D^2_L F_{\nu,{\rm thin}}}{R^3 c_5(p) \epsilon (p-2) E^{p-2}_{\rm min}} \left(\frac{e \gamma^2_m}{4 \pi m_e c c_1}\right)^{(p-1)/2}\right)^{1/3}\\ \nonumber 
 &= (1.6 \times 10^{-8}~{\rm G})~\left(\frac{F_{\nu,{\rm thin}}}{{\rm mJy}}\right)^{1/3} \left(\frac{R}{{\rm 100~kpc}}\right)^{-1/3} \left(\frac{D_L}{{\rm 100~Mpc}}\right)^{2/3},
\end{align}
assuming $\epsilon_e = 0.1$ and an index $p = 1-2 \alpha_{\rm low} = 2.94$ based on the inferred spectral index $\alpha_{\rm low} = -0.96$, and is indicative of emission from a weakened decelerating shock produced by Fermi acceleration of electrons \cite[e.g.][]{1987PhR...154....1B,1991SSRv...58..259J,2002apa..book.....F}. The estimated $\mu$G magnetic field strength is consistent with similar estimates for this source from other studies \citep{2021MNRAS.507..991S} and AGN in cluster environments \citep{2021MNRAS.tmp.2676M}. Assuming a spherical volume, the total energy content in the synchrotron emitting extended region is
\begin{equation}\label{E_ext}
E_{\rm ext} \approx \frac{4 \pi}{3} U_e R^3 = (1.82 \times 10^{56}~{\rm erg})~\epsilon ~\left(\frac{F_{\nu,{\rm thin}}}{{\rm mJy}}\right)^{2/3} \left(\frac{R}{{\rm 100~kpc}}\right)^{7/3} \left(\frac{D_L}{{\rm 100~Mpc}}\right)^{4/3},
\end{equation}
where, $U_e = \epsilon U_B = \epsilon B^2/(8 \pi)$, where $\epsilon = \epsilon_e/\epsilon_B$ and using $B$ from eqn. \ref{Blargescale}.
After eqn. \ref{E_ext} relating to the total energy $E_{\rm ext}$,
from the mean flux density measured in the MWA bands (88 -- 215 MHz), $\left< S_\nu ({\rm ext}) \right> = 0.246$ mJy, and using a spectral index $\alpha_{\rm low} = -1.15$ for the relics and a central frequency $\nu = 151.6$ MHz, the radio luminosity of the extended emission $L_{\rm R,ext} \approx 4 \pi D^2_L \nu \left< S_\nu ({\rm ext}) \right> (1+z)^{-1-\alpha_{\rm low}}= 6.18 \times 10^{36}$ erg s$^{-1}$. Using the empirical relations in eqn. \ref{Ljetrad} and eqn. \ref{Ljetrad1}, the associated luminosity of the extended jet is $L_{\rm ext} = 5.19 \times 10^{39}$ erg s$^{-1}$. 
 
\section{High frequency VLBI imaging of the pc-scale jet} \label{app:VLBIdata}

The source structure of \obj\ on parsec scales is revealed from the VLBI imaging data, which include the archive data from the Monitoring Of Jets in Active galactic nuclei with VLBA Experiments (MOJAVE) \citep{2018ApJS..234...12L} program and the archive data obtained from the Astrogeo database\footnote{\url{http://astrogeo.org/}}. Details of the VLBI data are presented in Table \ref{table1}.  The MOJAVE data of \obj\ were observed at 15 GHz over 25 sessions from 1995 July 28 to 2013 June 2. The Astrogeo data were observed simultaneously at the dual frequencies of 2.3 and 8.4 GHz, enabling the determination of spectral indices $\alpha_{\rm 2GHz}^{\rm 8GHz}$ for \obj. Since the source is unresolved in 2.3 and 8.4 GHz images, the analysis of the jet kinematics is based only on the 15 GHz VLBI data. 

All of these archival VLBI data have been calibrated, so we only made a few iterations of self-calibration in the {\sc Difmap} software package \citep{1995BAAS...27..903S} to eliminate some residual phase errors and to increase the dynamic range of the images.  Model fitting was performed using the MODELFIT program in {\sc Difmap}. \cite{2019ApJ...874...43L} studied the parsec-scale jet kinematics of 409 bright radio-bright AGN including \obj. They presented the model fitting results of \obj\ from epoch 1995 July 28 to epoch 2013 June 2. In their model fitting, (1) the source was detected with only a single core before 2011 January 11; (2) the distance of the fitted jet component from epoch 2000 July 22 to 2006 June 15 is less than 0.32 mas, i.e., less than half of the minor axis of the synthesized beam, therefore, these components are practically unresolvable. These above epochs were abandoned from the jet kinematics analysis. The model fitting parameters are given in Table \ref{table2}. 
The uncertainties of the parameters are first estimated from the equations given in \cite{1999ASPC..180..301F} which only take into account the fitting errors. In practical observations, the flux density error also includes a certain level of uncertainty in the flux scale calibration through error propagation. For the 15-GHz VLBA, this systematic error is typically about 5\%. Both total intensity (Stokes I) and linear polarization images are created from the MOJAVE data. After self-calibration in {\sc Difmap}, we separately created images using the Stokes I, Q and U data. Then we combined the Stokes Q and U images to produce the linear polarization intensity $p$, which is the root of the sum of squares of Q and U components, i.e., $p = \sqrt{Q^2 + U^2}$. The fractional polarization was calculated as the ratio of the polarized flux density to Stokes I flux density, $p/I$. The derived fractional polarizations are consistent with the values reported on the MOJAVE webpage.
Snapshot-mode Astrogeo data have short integration time and sparse (u,v) coverage. Deconvolution and model fitting are affected by strong sidelobes caused by incomplete (u,v) coverage. For this reason, the uncertainty in component size adopts the root of the quadratic sum of its corresponding statistical error and the fitting error. 
The position error was then estimated as half of the component size error.

The present paper is focused on the 2009--2010 major flare and its associated jet properties, so we only use the MOJAVE data after 2004, including 13 epochs listed in Table \ref{table1}. The (\textit{u,v}) coverage, image sensitivity, and resolution of these VLBI data are all in good agreement, therefore the systematic errors between the fitted parameters are small.
In six of the 13 epochs, the jet is not distinguishable, and these data were not used in jet kinematics analysis. 

On pc scales, VLBI images of \obj\ reveal either a single core or ``a core + a compact jet" structure. The compact core dominates the total flux density in all images. The jet extends to the west and is relatively weaker but well distinguished from the core in 7 epochs, allowing us to determine its kinematic properties. 
We double-checked the model fitting reported by the MOJAVE team and only adopted the high-confidence jet models. If the signal-to-noise ratio of a jet component is lower than 3, or the jet is indistinguishable from the core, then it will not be used for the kinematic analysis. 
Figure \ref{fig:VLBI-2} shows the variation of the core-jet separation with time and the variation of the jet position angle with time. It is clear that these jet knots do not belong to a single component, but are two separated jet knots, labeled J1 and J2 in the plot (corresponding to components \#1 and \#4  in \citealt{2019ApJ...874...43L}), each of which follows a ballistic trajectory along a different position angle. 
We made linear regression fitting to the position-time correlation of J1 and J2 and obtained the jet proper motions as $0.24 \pm 0.02$ mas yr$^{-1}$ and $0.22 \pm 0.01$ mas yr$^{-1}$, respectively. These convert to jet transverse speeds of $1.35 \pm 0.13\, c$ (J1) and $1.21 \pm 0.07 \,c$ (J2). The derived jet proper motions are in good agreement with the previous studies \citep{2000A&A...357L..45B} based on the 43-GHz VLBI observations in 1998.
To distinguish it from our jet components, we name the jet detected by Brunthaler et al. as ``1998 jet". Extending the 1998 jet to epoch 2006 (the first epoch of the MOJAVE data used in this paper), we find the component should be at a distance of about 0.525 mas. However, it is not detected in the 2006 VLBA image, suggesting that this component has been considerably dimmed due to the adiabatic radiation loss.

\begin{table*}
    \centering
    \caption{Logs of the 15 GHz VLBI observations.}
    \begin{tabular}{cccccccc}
    \hline \hline 
    Date         & Project Code & Bandwidth  & On-source time &Beam                             & $S_{\rm peak}$  & $\sigma_{\rm rms}$  \\
    (yyyy-mm-dd) &              & (MHz)      &(hour)           & (maj, min, PA) & (mJy beam$^{-1}$)  &(mJy beam$^{-1}$)     \\
    (1) & (2) & (3) & (4) & (5) & (6) & (7) \\
    \hline 
    
    2004-02-11   &BL111        &32.0         &1.05            & $1.2\times0.6$ mas$^2$, $-4\fdg1$   & 1363.9   & 0.19      \\   
    2005-03-05   &BL123        &32.0         &0.93            & $1.1\times0.5$ mas$^2$, $-4\fdg3$   & 1680.1   & 0.29      \\   
    2005-05-26   &BL123        &32.0         &0.93            & $1.1\times0.6$ mas$^2$, $-6\fdg4$   & 1755.0   & 0.31      \\     
    2005-12-22   &BL123        &32.0         &0.82            & $1.2\times0.6$ mas$^2$, $-5\fdg1$   & 1083.9   & 0.24      \\    
    2006-06-15   &BL137        &32.0         &0.93            & $1.2\times0.6$ mas$^2$, $-7\fdg3$   &  725.6   & 0.37      \\
    2007-08-09   &BL149        &32.0         &0.93            & $1.2\times0.6$ mas$^2$, $-4\fdg4$   &  293.9   & 0.22      \\
    2008-08-25   &BL149        &32.0         &1.05            & $1.2\times0.5$ mas$^2$, $-8\fdg1$   &  482.9   & 0.24      \\
    2009-06-03   &BL149        &64.0         &1.05            & $1.2\times0.6$ mas$^2$, $-5\fdg2$   & 1204.1   & 0.16      \\
    2010-07-12   &BL149CL      &64.0         &1.05            & $1.4\times0.5$ mas$^2$, $-7\fdg7$   & 1138.9   & 0.18      \\  
    2011-05-26   &BL149DI      &64.0         &1.17            & $1.2\times0.6$ mas$^2$, $-3\fdg5$   &  136.4   & 0.18      \\
    2012-04-30   &BL178AJ      &64.0         &1.28            & $1.3\times0.6$ mas$^2$, $-6\fdg6$   &  140.3   & 0.17      \\
    2012-11-02   &BL178AR      &64.0         &0.93            & $1.2\times0.6$ mas$^2$, $-4\fdg9$   &  622.7   & 0.15      \\
    2013-06-02   &BL178BD      &64.0         &1.05            & $1.2\times0.6$ mas$^2$, $0\fdg1$    &  557.4   & 0.19      \\     
    
    \hline
    \end{tabular} \\
    Note: Column 5 gives the restoring beam in natural weighting; Columns 6 and 7 present the peak flux density and {\it rms} noise in the images, respectively. 
    \label{table1}
\end{table*}

\begin{table*}
    \centering
    \caption{Model fitting results of 15-GHz VLBI data.}
    \begin{tabular}{cccccccc}
    \hline \hline 
    Time         & Comp. & $S_{\rm int}$ & R                  & PA               &$\theta_{\rm FWHM}$ &$T_{\rm B}$   & $\delta$\\
    (yyyy-mm-dd) &       & (mJy)         &(mas)               &($\degr$)             &(mas)               &(10$^{10}$ K) &  \\ 
    (1) & (2) & (3) & (4) & (5) & (6) & (7) & (8) \\
    \hline 
    2004-02-11   &C      &1390$\pm$52     &0                  &0                  &0.098$\pm$0.001     & 79.6$\pm$3.00 & 15.9  \\
    2005-03-05   &C      &1855$\pm$186    &0                  &0                  &0.225$\pm$0.003     & 18.6$\pm$2.05 &  3.7  \\
    2005-05-26   &C      &1890$\pm$134    &0                  &0                  &0.203$\pm$0.001     & 23.9$\pm$1.82 &  4.8  \\
    2005-12-22   &C      &1358$\pm$156    &0                  &0                  &0.380$\pm$0.007     & 4.25$\pm$0.61 &  0.9  \\
    2006-06-15   &C      &642$\pm$120     &0                  &0                  &0.175$\pm$0.005     & 13.5$\pm$2.2  &  2.7  \\
                 &J1     &291$\pm$9       &0.417$\pm$0.004    &$-67.2\pm0.5$      &0.263$\pm$0.008     & 0.42$\pm$0.07 &  ...  \\
    2007-08-09   &C      &296$\pm$58      &0                   &0                 &0.120$\pm$0.004     & 11.9$\pm$2.3  &  2.4  \\
                 &J1     &81$\pm$8        &0.701$\pm$0.006     &$-70.5\pm0.5$     &0.437$\pm$0.012     & 0.15$\pm$0.02 &  ...  \\
    2008-08-25   &C      &488$\pm$24      &0                   &0                 &0.072$\pm$0.002     & 52.6$\pm$2.6  & 10.5  \\
                 &J1     &20$\pm$3        &1.035$\pm$0.006     &$-67.8\pm0.3$     &0.416$\pm$0.012     & 0.04$\pm$0.01 &  ...  \\
    2009-06-03   &C      &1219$\pm$44     &0                   &0                 &0.084$\pm$0.001     & 95.9$\pm$3.4  & 19.2  \\
                 &J1     &6.9$\pm$1.5     &1.123$\pm$0.013     &$-68.1\pm0.7$     &0.350$\pm$0.026     & 0.03$\pm$0.01 &  ...  \\
    2010-07-12   &C      &1203$\pm$74     &0                   &0                 &0.173$\pm$0.001     & 21.3$\pm$1.4  &  4.3  \\
    2011-05-26   &C      &121$\pm$22      &0                   &0                 &0.147$\pm$0.003     & 3.58$\pm$1.38 &  0.7  \\
                 &J2     &46$\pm$2        &0.390$\pm$0.004     &$-63.2\pm0.6$     &0.177$\pm$0.008     & 0.16$\pm$0.03 &  ...  \\
    2012-04-30   &C      &141$\pm$15      &0                   &0                 &0.099$\pm$0.007     & 8.11$\pm$0.86 &  1.6  \\
                 &J2     &11$\pm$1        &0.583$\pm$0.004     &$-63.2\pm0.4$     &0.328$\pm$0.008     & 0.33$\pm$0.01 &  ...  \\
    2012-11-02   &C      &625$\pm$18      &0                   &0                 &0.048$\pm$0.001     & 151.9$\pm$4.4 & 30.4  \\
                 &J2     &3.9$\pm$0.8     &0.714$\pm$0.029     &$-62.6\pm2.3$     &0.606$\pm$0.058     & 0.01$\pm$0.01 &  ...  \\
    2013-06-02   &C      &565$\pm$31      &0                   &0                 &0.086$\pm$0.001     & 42.3$\pm$2.4  &  8.5  \\       
    \hline
   \end{tabular}\\
Columns (3) to (6) present the model fitting parameters in sequence: the integrated flux density, radial separation, position angle with respect to the core C (measured from north through east), component size (full width at half maximum of the fitted Gaussian component) and brightness temperature. Note that the uncertainty of the position (R) and component size ($\theta_{\rm FWHM}$) only takes into account of the statistical error, which is inverse proportional to the signal-to-noise ratio of the component.
   \label{table2}
\end{table*}

\section{Flux density variability and time series analysis}\label{app:singledish}

The single-dish light curve data used in this paper were from the archives of the 14-meter Mets\"{a}hovi radio telescope in Finland and the 40-meter telescope of the Owens Valley Radio Observatory (OVRO) in US.

Mets\"{a}hovi observations of \obj\ were made at 22 and 37 GHz in the period of 1986--2020. The 22- and 37-GHz light curves from 1984 to 1998 have been published in \citet{1999ApJ...514L..17F}, and the Mets\"{a}hovi data from 1986 to 2019 have been published in \citet{2020A&A...635A.172C}. In addition to these data, we also include the new data from 2019 onwards to date. 
A detailed description of the data reduction and analysis is given in \citet{1998A&AS..132..305T}. Under good observing conditions, the detection limit of the Mets\"{a}hovi telescope at 37 GHz is around 0.2 Jy. Uncertainties in flux densities include the contribution from the root mean square (\textit{rms}) of the measurements and the errors in the absolute flux density calibration.  

The radio monitoring program with the 40 m telescope at the OVRO includes over 1500 sources in the northern sky ($\delta > -20^\circ$) \citep{2011ApJS..194...29R}. Each source is observed twice per week. The minimum flux density detected is about 4 mJy. A typical uncertainty of the measurements is $\sim$3\%. The high cadence and high sensitivity of the monitoring greatly facilitate the study of the variability properties of radio sources on timescales of months and years. The OVRO monitoring data of \obj\ were observed at 15 GHz from 2008 to 2020. The maximum and minimum flux densities are 1825 mJy and 72 mJy, respectively. The integrated flux densities from the 15 GHz VLBA data are superimposed on the OVRO light curve, showing a good match between the two sets of flux densities at adjacent epochs.  This suggests that the steep-spectrum extended structures including the kpc-scale jet and lobes seen in the low-frequency images are barely detectable at 15 GHz and above.

The time series analysis is carried out using two methods, the Fourier periodogram  \citep{1989ApJ...343..874S,2005A&A...431..391V,2013MNRAS.434.3487A,2014ApJ...791...74M} and the wavelet analysis  \citep{1998BAMS...79...61T,2013MNRAS.434.3487A,2016MNRAS.456..654M}. For both methods, the light curves are first pre-processed by a conversion from a partial uneven sampling with small data gaps 
to a full even sampling after a linear interpolation and re-sampling. The normalized Fourier periodogram $P (f_j)$ is evaluated as \citep{2005A&A...431..391V,2014ApJ...791...74M,2015MNRAS.452.2004M} 
\begin{equation}
P (f_j) = \frac{2 \Delta t}{\overline{x}^2 N} |F(f_j)|^2,
\end{equation}
where $\Delta t$ is the sampling time step size, $\overline{x}$ is the mean of the light curve $x(n \Delta t)$ of length $N$, and $F(f_j)$ is the Fourier transform of $x$ evaluated at the frequencies $f_j = j/(N \Delta t)$ with $j = 1, ....., (N/2-1)$ (up to the Nyquist frequency). Since astrophysical processes typically produce red noise in the light curve, the periodogram $P(f_j)$ will be characterized by a power law behavior, especially at low temporal frequencies. This can be captured by parametric model fits to $P(f_j)$ to estimate the underlying power spectral density (PSD). Here, we use two models, a power law given by
\begin{equation}\label{model:plaw}
I(f_j) = A f^{\alpha}_j+C,
\end{equation}
where $A$ is the normalized amplitude, $\alpha$ is the power-law index, $C$ is the ambient noise level, and a bending power law is given by
\begin{equation}
I(f_j) = A f^{-\alpha_l}_j (1+(f_j/f_b)^{\alpha-\alpha_l})^{-1}+C,
\end{equation}
where $f_b$ is a break frequency marking a transition in slopes, the power-law indices are $\alpha$ for $f_j > f_b$, and $\alpha_l$ for $f_j < f_b$, and $C$ is the ambient noise level. 

The fitting of the Fourier periodogram with the above parametric models is carried out using the maximum likelihood estimator. The methodology and estimation of parameters and their errors are discussed in \citep{2014ApJ...791...74M,2015MNRAS.452.2004M}. After accounting for model uncertainties, statistical significance testing is carried out to identify outlying peaks in the fit residuals (based on an expected $\chi^2$ statistical distribution), which are potential candidates for quasi-periodic oscillations in the light curve. 

The statistical significance of all peaks is assessed based on a procedure involving Monte Carlo simulations  \citep{2014ApJ...791...74M,2015MNRAS.452.2004M}. 
These are carried out using the algorithm of \citet{1995A&A...300..707T}. The best-fit model and associated parameters are used as trial values to simulate 5000 realizations of the periodogram, oversampled in duration and temporal frequencies, and then re-sampled at the original frequencies.
A mean periodogram $\tilde{I}(f_j)$ is constructed from the simulations and is re-scaled to match the variability properties of the original periodogram; this could be the closest estimate of the underlying PSD. The statistical significance of the periodogram ordinates for any frequency bin is evaluated based on the assumption that the light curve consists of randomly distributed data points (i.e., no periodic behavior). The residuals from the fitting, $P(f_j)/\tilde{I}(f_j)$, are then $\chi^2_2/2$ distributed \citep{chatfield2016analysis}, with a conditional probability $p [P(f_j)|\tilde{I}(f_j)] = \tilde{I}(f_j)^{-1} e^{-P(f_j)/\tilde{I}(f_j)}$. The cumulative distribution function for the PSD ordinates is then the integral of the $\chi^2_2$ distribution, i.e., a gamma density function $\Gamma(1,1/2)= \exp{(-x/2)}/2$. Specifying a level of statistical significance $(1-\epsilon)$ in the integral helps to identify outliers (quasi-periodic signals) that may be present in the tail of the distribution. 
We set a threshold of $\epsilon = 0.05$ (95 \% level of significance) to identify quasi-periodic signals in the light curves.

The wavelet analysis is employed here in a complementary manner to probe quasi-periodic signals in the light curves and their locations (to infer the total duration and number of cycles). The two-dimensional wavelet power spectrum is a function of the wavelet scale (that can be expressed in units of the sampling wavelength or period) and the time window being sampled, and is evaluated as \citep{2016MNRAS.456..654M}
\begin{align}
P (n,s) &= |W(n,s)|^2 \\ \nonumber
W (n,s) &= \sum^{N}_{j=1} F(2 \pi f_j) \Psi^{\ast} (2 \pi s f_j) e^{2 \pi i f_j n \Delta t},
\end{align}
where $W(n,s)$ is the wavelet transform of the evenly sampled light curve $x (n \Delta t)$ at times $(n \Delta t)$ and at scales $s$, $F(2 \pi f_j)$ is the Fourier transform of $x$ evaluated at the circular frequencies $2 \pi f_j = 2 \pi j/(N \Delta t)$ with $j = 1, ....., N$, and $\Psi^\ast (2 \pi s f_j)$ is the complex conjugate of the Fourier transform of the wavelet sampling kernel function. The wavelet transform is the inverse Fourier transform of the convolution product of the above constituents. For a continuous wavelet transform, a commonly used sampling kernel is the Morlet wavelet function \citep{1989wtfm.conf....2G} $\psi = \pi^{-1/4} e^{i \omega_0 t} e^{-t^2/2}$ where $\omega_0 = 6$ is a frequency parameter characterizing the wavelet shape and $t$ is the time parameter. In the frequency domain $\Psi (2 \pi s f_j) = \pi^{-1/4} e^{-(2 \pi s f_j-\omega_0)^2/2}$. The wavelet scales $s \approx 1.03/f_j$ for the Morlet function  \citep{1998BAMS...79...61T} is hence in near correspondence with the sampling frequencies. In the current work, the following additional features have been implemented: the use of a cone of influence to help explain the cyclic behavior of the sampling kernel, especially at low temporal frequencies \citep{1998BAMS...79...61T}, the use of a Hann window function to smoothen the noisy features in the power spectrum, and the identification of contiguous features in the power spectrum that may be statistically significant in anticipation of measuring their total duration, number of cycles and the time evolution of the features. These improvements all tend to narrow the search window and consequent computational costs in the identification of statistically significant signals, and considerably improve the contrast between the signal and noise.

The statistical significance of contiguous signals detected in the wavelet analysis employs a two-fold strategy. In the first stage, the algorithm of \citet{1995A&A...300..707T} and the expected $\chi^2_2$ statistics of periodogram ordinates (assuming that the light curve ideally consists of random Gaussian noise) are employed to simulate a large number of realizations of the periodogram. The best-fit power-law model of the form $I(f_m) = A f^{\alpha}_m+C$ with associated parameters is used as trial values. The index $\alpha$ is varied in the range from the best-fit value to 0.0 in steps of $-0.2$ and in each case. 1000 realizations of the periodogram are simulated, and  they are oversampled in duration and temporal frequencies and then re-sampled at the original frequencies. In each realization, the periodogram is inverse Fourier transformed to obtain a synthetic light curve with similar statistical and variability properties as that of the original light curve; the total number of simulated light curves is typically $\geqslant 20000$. In the second stage, their individual wavelet power spectra are evaluated. For each simulated wavelet power spectrum, the mean of all powers at a given scale $P(n)$ is used to estimate the global wavelet power spectrum GWPS$(s)$ that corresponds to a window function smoothed version of the Fourier periodogram. The candidate signals in the GWPS of the original light curve are compared with their simulated counterparts. The number of times $p$ that a candidate signal in the original GWPS exceeds the values in the simulations (totaling $Q$) at a given wavelet scale is measured in terms of the statistical significance $(1-p/Q)$.

\section{Integrated radio spectrum and fitting}\label{app:spectrum}

We plot the radio spectrum of \obj\ from 72 MHz to 37 GHz in Figure \ref{fig:SED}. The red-colored squares below 230 MHz are taken from the MWA GLEAM-X survey \citep{2022PASA...39...35H} and the green-colored diamond is from the GMRT observation \citep{2017A&A...598A..78I}, revealing the extended radio emission characterized by a steep spectrum. 
The data point in slateblue color is from the ASKAP observation at 888 MHz. The solid circle in magenta color is from the recent uGMRT observation \citep{2020MNRAS.499.5826S} on 2018 November 23. 
In addition to these low-frequency data, we also include the data points at GHz frequencies observed in the epochs close to the MWA and uGMRT observation: 37 GHz Mets\"{a}hovi (blue-colored triangle-left), 15 GHz OVRO (navy-colored triangle-right), 8.4 and 2.3 GHz Astrogeo VLBI (purple-colored diamond). From Figure \ref{fig:lightcurve} we find that the core dominates the total flux density at GHz frequencies. 

The spectrum shows different characteristics above and below 685 MHz: at 685 MHz and above, the core dominates the emission, showing an inverted spectrum; at frequencies below 685 MHz, the flux density from the core is substantially reduced toward the low-frequency end due to increasing synchrotron self-absorption, and the extended jets and lobes become increasingly dominant. The two-component radio spectrum spanning  0.072 GHz -- 37 GHz is subjected to a weighted (the measurement errors of flux densities are used as weights) least square fit using the function eqn. \ref{Fnufit}, in which the first two parts ($A \nu^{\alpha_{\rm low}}+B$) form a power-law spectrum and the last one describes a spectrum of self-absorbed synchrotron radiation 
\citep{1970ranp.book.....P,1999A&A...349...45T}.
\begin{equation}
    F_\nu = A \nu^{\alpha_{\rm low}}+B+F_m \left(\frac{\nu}{\nu_m}\right)^{\alpha_{\rm thick}} \left\{ \frac{1-e^{-\tau_m (\nu/\nu_m)^{\alpha-\alpha_{\rm thick}}}}{1-e^{-\tau_m}}\right\} \\ \nonumber
\end{equation} 
\begin{equation}
    \tau_m = \frac{3}{2} \left(\left(1-\frac{8 \alpha}{\alpha_{\rm thick}}\right)^{1/2}-1\right),
    \label{Fnufit}
\end{equation}
where the parameters of the low-frequency power-law spectrum include the amplitude $A$ (mJy), the spectral index $\alpha_{\rm low}$, a baseline flux density $B$ (mJy), and the parameters of the high-frequency section include the amplitude $F_m$ (mJy), the frequency of transition from optically thick to thin emission $\nu_m$ (GHz),  the optically-thick spectral index $\alpha_{\rm thick}$, the optically-thin spectral index $\alpha$, and the optical depth $\tau_m$ expressed in terms of $\alpha_{\rm thick}$ and $\alpha$.
In the fitting process  derived by using the Markov chain Monte Carlo (MCMC) method, $\alpha_{\rm thick}$ is  fixed at 2.5, corresponding to the canonical synchrotron self-absorption case, and other parameters are constrained within reasonable ranges $0.0 < A \leq 0.8$ Jy, $-1.5 \leq \alpha_{\rm low} \leq -0.4$, $0.0 < B \leq 0.6$ , $0.0 < F_m \leq 0.5$ Jy, $7.0 \leq \nu_m \leq 15.0$ GHz, and $-0.4 \leq \alpha \leq 0.0$. This yields $A = 27^{+12}_{-9} $ mJy, $\alpha_{\rm low} = -0.97^{+0.21}_{-0.21}$, $B = 41^{+8}_{-12}$ mJy, $F_m = 146^{+11}_{-9}$ mJy, $\nu_m = 11.24^{+2.01}_{-1.25}$ GHz, and $\alpha = -0.20^{+0.10}_{-0.12}$.

In Figure \ref{fig:SED} we also plotted the flux densities of the NE and SW lobes obtained from the VLA images \citep{2005A&A...435..497B} and fitted the NE and SW lobes  (represented by blue-colored open circles and red-colored open squares, respectively) with power law functions. We then calculated the flux densities of NE and SW lobes at the MWA observation frequencies (i.e., 88, 118, 154, and 200 MHz) by extrapolating the power-law fits. Next, we subtracted the extrapolated flux densities of NE and SW lobes from the measured values to estimate the flux densities from large-scale extended emission contributed mainly by the N and S lobes. Finally, we calculated the spectral index of the extended outer lobes N+S as $\alpha = -1.09 \pm 0.12 $, which is much steeper than the inner lobes and comparable to the spectral index of the recently observed radio relics at low frequencies, indicating that the outer lobes are more likely to be dying relics, dominated by radiation from aged relativistic electrons.

\section{Emission properties of the pc -- kpc scale jet} \label{app:emissionproperties}

The synchrotron emission spectrum is assumed to be shaped by a power-law energy distribution of the emitting relativistic electrons $N (E)~dE = K E^{-p}~dE$ for energy $E$, normalization $K$ and energy index $p \geqslant 2$. The normalization and magnetic field strength $B$ are expressed in terms of the radial distance $r$ from the supermassive black hole as $K = K_0 (r/r_0)^{-2}$ and $B = B_0 (r/r_0)^{-1}$ with scaling constants $K_0$ and $B_0$ at a fiducial distance $r_0$  \citep{2015MNRAS.451..927Z,2015MNRAS.452.2004M,2017MNRAS.469..813A}. With this, the particle kinetic energy density is
\begin{equation}
U_e = \epsilon_e \int^{\infty}_{E_{\rm min}} N (E) E~dE = \epsilon_e K_0 \left(\frac{r}{r_0}\right)^{-2} \frac{E^{-p+2}_{\rm min}}{(p-2)},
\end{equation}
where $\epsilon_e$ is the fraction of the total energy density in the particle kinetic energy and $E_{\rm min}$ is the minimum energy required to accelerate electrons (assuming that they are the dominant constituents of the jet) to relativistic energies, here taken to be the rest mass energy $E_{\rm min} = 0.51$ MeV. Assuming equipartition of the total energy density between that in the magnetic fields $U_B = B^2/(8 \pi)$ and in the particle kinetic energy $U_e$, i.e. $U_e/\epsilon_e = U_B/\epsilon_B$, the normalization constant is
\begin{equation}\label{K0eqn}
K_0 = \frac{\epsilon B^2_0}{8 \pi} (p-2) E^{p-2}_{\rm min},
\end{equation}
where $\epsilon_B$ is the fraction of the total energy density in the magnetic fields and $\epsilon = \epsilon_e/\epsilon_B$. The total energy density is then
\begin{equation}
U_{\rm tot} = U_e+U_B = U_B (1+\epsilon) = \frac{B^2_0}{8 \pi} \left(\frac{r}{r_0}\right)^{-2} (1+\epsilon).
\end{equation}
 The jet luminosity attributable to synchrotron emission in a region of size $\varpi = r \sin \theta_j$ (where $\theta_j$ is the jet half opening angle) downstream of the jet base is  \citep{2010MNRAS.409L..79G}
 \begin{equation}\label{Ljet}
 L_{\rm jet} = \pi \varpi^2 \beta_j \Gamma^2_j c U_{\rm tot} = \frac{\beta_j \Gamma^2_j c}{8} \sin^2 \theta_j (B_0 r_0)^2 (1+\epsilon),
 \end{equation}
where $\beta_j$ and $\Gamma_j$ are the bulk velocity and Lorentz factor of the jet. The optically thick absorption coefficient \citep{1986rpa..book.....R,2013LNP...873.....G}
\begin{equation}\label{alphanu}
\alpha_{\nu'} = \frac{\pi^{1/2} e^2 K}{8 m_e c} \left(\frac{e B}{2 \pi m_e c}\right)^{(p+2)/2} (\nu')^{-(p+4)/2} f(p),
\end{equation}
where $e$ is the electron charge, $\nu'$ is the emission frequency in the source rest frame, and $f(p)$ is expressed in terms of $p$ and the Gamma function $\Gamma$ dependence as
\begin{equation}
f(p) = \frac{3^{(p+1)/2} }{\displaystyle \Gamma \left(\frac{p+8}{12}\right)} \Gamma \left(\frac{3 p+22}{12}\right) \Gamma \left(\frac{3 p+2}{12}\right) \Gamma \left(\frac{p+6}{12}\right).
\end{equation}
The optical depth for the emitting region $\varpi$ is $\tau_{\nu'} = \alpha_{\nu'} \varpi$. From the condition $\tau_{\nu'} = 1$ that signifies the transition of the emitting region from optically thick to thin, the associated radial distance corresponds to the location of the emitting core. The associated synchrotron self-absorption frequency $\nu'  = \nu (1+z)/\delta$ where $\nu$ is the frequency in the observer frame, accounting for cosmological redshift through the factor $(1+z)$ for a source at redshift $z$, and relativistic beaming through the Doppler factor $\delta$. Using eqns. \ref{K0eqn}, \ref{Ljet} and \ref{alphanu} in the above condition,
\begin{equation}\label{rSSA}
r = \frac{\nu^{-1} \delta}{(1+z)} \left(\frac{\pi^{1/2} e^2 \epsilon E^{(p-2)}_{\rm min} \sin \theta_j (p-2) f(p)}{64 \pi m_e c}\right)^{\frac{2}{(p+4)}} \left(\frac{e}{2 \pi m_e c}\right)^{\frac{(p+2)}{(p+4)}} \left(\frac{8 L_{\rm jet}}{(1+\epsilon) \sin^2 \theta_j \beta_j \Gamma^2_j c}\right)^{\frac{(p+6)}{2 (p+4)}},
\end{equation}
with the familiar $r \propto \nu^{-1}$ as expected for a self-absorbed radio core  \citep{1995A&A...293..665F,1998A&A...330...79L}. 

The mean flux density of the radio core at 15 GHz from the VLBI data is $\left< S_\nu (C) \right> = 907.15 \pm 90.87$ mJy (see Table \ref{table2}). The associated radio luminosity $L_R = 4 \pi \nu D^2_L \left< S_\nu (C) \right> (1+z)^{-1+\alpha} = 2.37 \times 10^{42}$ erg s$^{-1}$ for $\nu = 15$ GHz, $D_L = 403.1$ Mpc, and $z = 0.089$. We use empirical relations to estimate the total jet luminosity (radiative $L_{\rm jet,rad}$ and kinetic $L_{\rm jet,kin}$ components, associated with emission and the acceleration of baryonic constituents of the jet, respectively) from the radio luminosity  \citep{2014IJMPS..2860188F,2020NatCo..11..143A}
\begin{equation}\label{Ljetrad}
    \log L_{\rm jet,rad} = 12.00+0.75 \log L_{R}
\end{equation} 
\begin{equation}\label{Ljetrad1}
    \log L_{\rm jet,kin} = 6.00+0.90 \log L_{R}.
\end{equation} 
The total jet luminosity $L_{\rm jet} =  L_{\rm jet,rad} +L_{\rm jet,kin}  = 1.97 \times 10^{44}$ erg s$^{-1}$. 

Jet properties are estimated from the VLBI 15 GHz data points during the flaring phases ($\delta \geqslant$ 4.3; median Doppler factor). These include an average $\tilde{\delta} = 16.9$, an associated $\Gamma = 8.5$ based on $\Gamma = (\beta^2_{\rm app}+\tilde{\delta}^2+1)/(2 \tilde{\delta})$. The jet half opening angle $\theta_j = 1/\Gamma = 6\fdg7$, similar to the lower limit estimated in \citet{2021A&A...652A..14C}. With the estimated $L_{\rm jet}$, $p = 2.3$, the assumption of equipartition in the energy density between that in the magnetic fields and particle kinetic energy with $\epsilon = 1$ and the above physical properties, the radial distance of the self-absorbed core from eqn. \ref{rSSA}, $r_{\rm SSA} \approx 435 r_G$ (where $r_G = G M_\bullet/c^2 = 1.47 \times 10^{13}$ cm is the gravitational radius for a SMBH of $1.84 \times 10^8~M_\odot$ \citep{2012ApJ...755...60G}). Using this in eqn. \ref{Ljet} ($r_0 = r_{\rm SSA}$), the associated magnetic field strength is $B_0 = 13.8$ G. For the $B \propto r^{-1}$ scaling relation, this corresponds to $B (r = 1~{\rm pc}) = 52.8$ mG. This estimate is consistent with the core-shift based measurement of $\leqslant 60$ mG and of a similar order of magnitude to the SSA-based measurement of $\leqslant 20$ mG  \citep{2021A&A...652A..14C}.

\section{Decomposition of the major flaring phase during 2009--2010 and origin} \label{app:2009flare}

\begin{figure}
    \centering
    \includegraphics[width=0.45\textwidth,height=7cm]{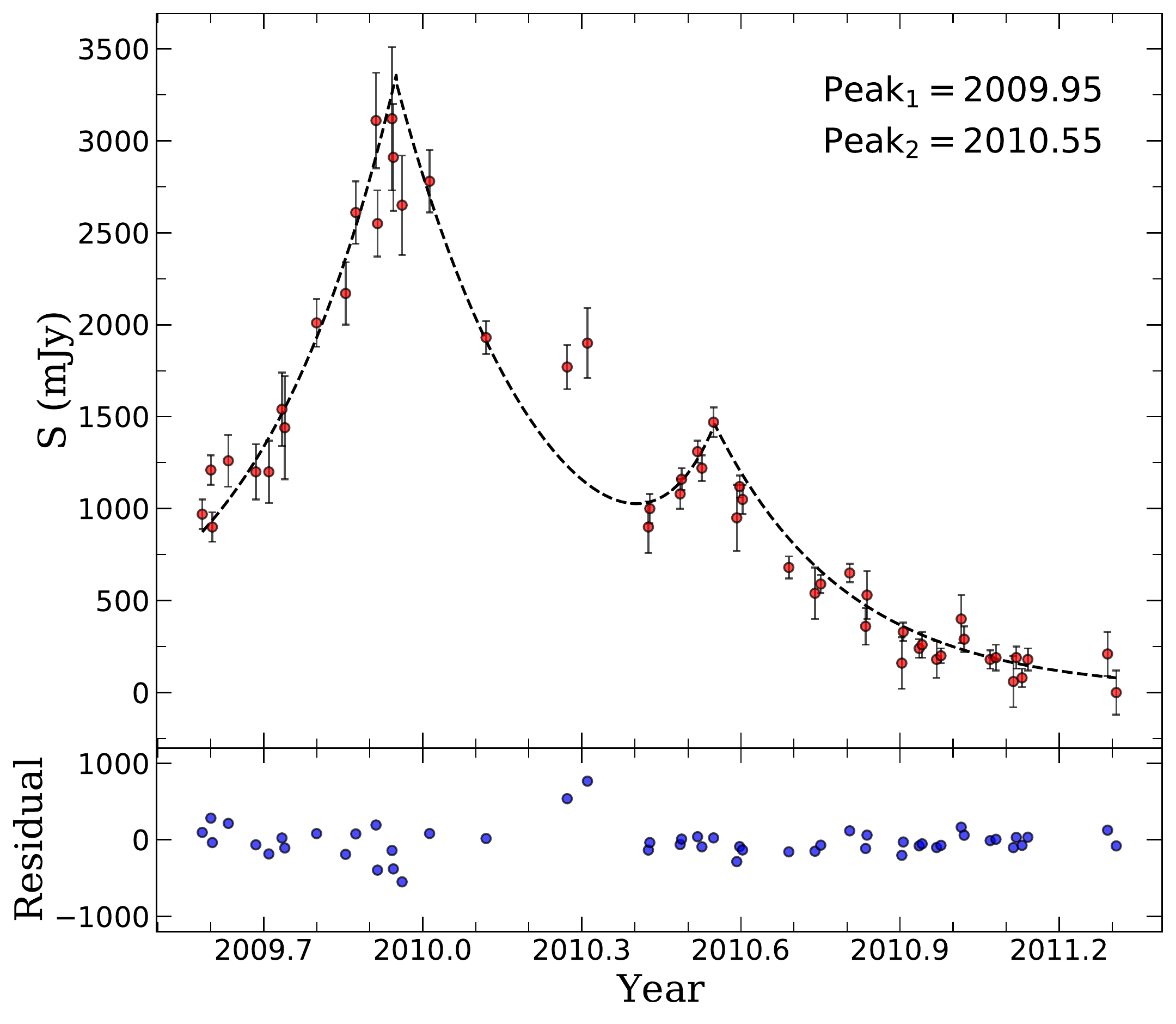}    
    \includegraphics[width=0.45\textwidth,height=7cm]{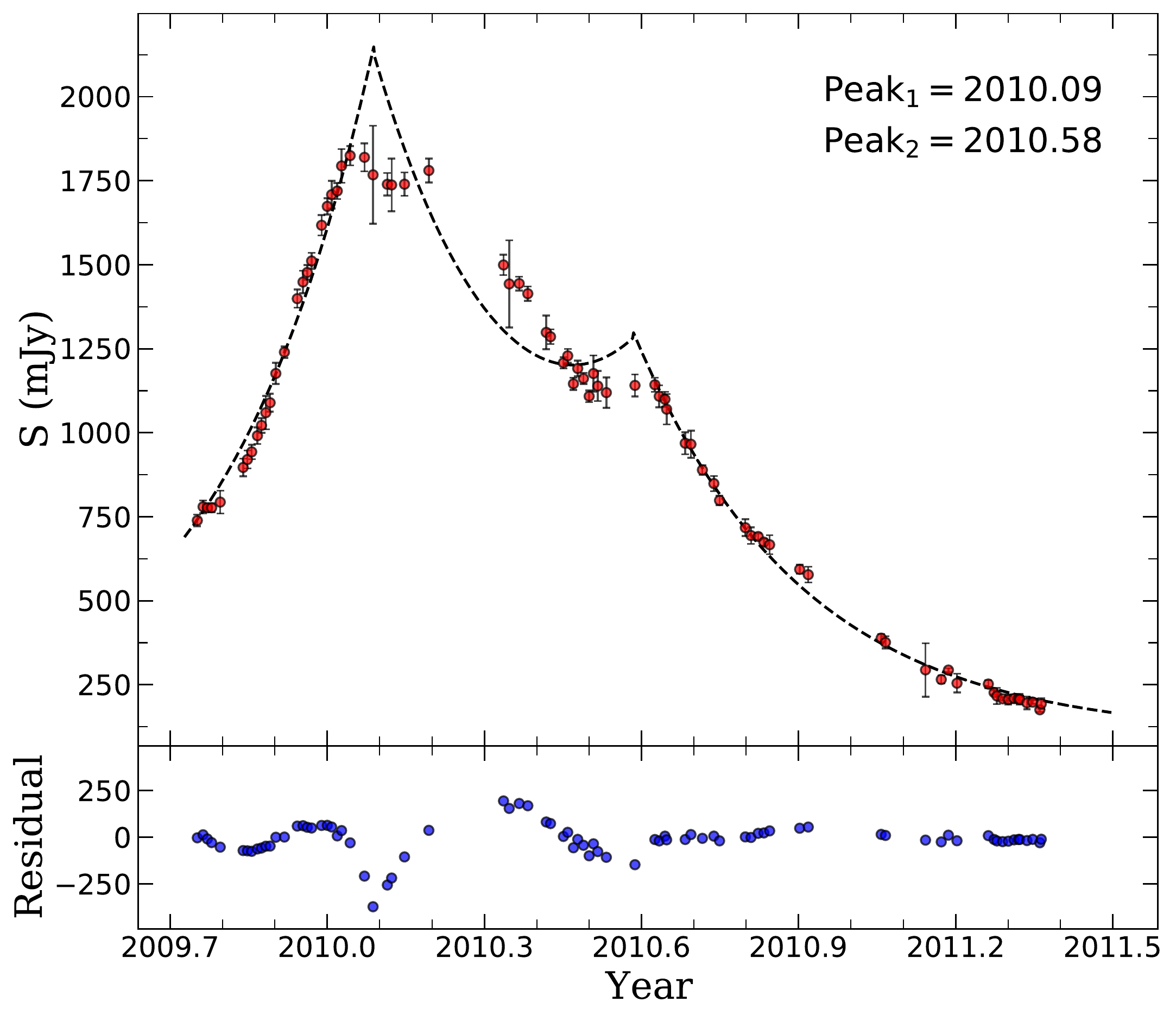}
    \caption{Decomposition of the major flare in 2009--2010 with two exponential components. {\it Left}: the 37 GHz light curve; {\it Right}: the 15 GHz lightcurve. The uncertainties of the fitted parameters only take into account the statistical error. }
    \label{fig:2009flare}
\end{figure}

The AGN flares based on generalized shock models  can be described by a fast-rising and slowly-declining pattern \citep{1999ApJS..120...95V}, which can be modeled with exponential functions. However, the rising phase of the flux of \obj\ does not exhibit a significantly shorter timescale than the declining phase. For example, the rise of the 1998--1999 flare of \obj\ is even slower than the decay timescale \citep{2005A&A...435..497B}. 
Several lower-amplitude flares were observed prior to the 2009--2010 flare peak, similar to the 1999 and 2004 flares. On the one hand, these smaller flares did not form jets observable in VLBI images, and on the other hand, the opacity of the core and the limited imaging resolution hindered the detection of the corresponding fine structural changes.
However, the superposition of these sub-flares leads to a slow increase in flux density before the peak of the major flare. 
Therefore, it is difficult to obtain a 1:1.3 ratio (as found in many other blazars by \citealt{1999ApJS..120...95V}) between the rising and declining timescales in \obj.
For this reason, we did not fix the decay-to-rise ratio in exponential functions used to model the \obj\ flares.

In addition, two $\gamma$-ray flares were found during the 2009-2010 flare, one is associated with the peak of 2009 flare, and the other is in mid of 2010.
This motivates us to decompose the light curves with two flare components as an approximate description of the primary 2009 flare and the subsequent 2010 flare (Figure \ref{fig:2009flare}), respectively. 
We need to mention that the model curves shown in Figure \ref{fig:2009flare} are not obtained by least-squares fitting to the observed data, but by selecting model curves with minimum residuals from those obtained with different parameter combinations.

In the main text, we inferred a correlation between $\gamma$-ray and prominent radio bursts, and that the flares are directly connected to the production of jet knots. These can be explained by the shock-in-jet model developed for the blazars \citep{2008Natur.452..966M}. Then where did these events occur ?

From Figure \ref{fig:VLBI-2} we find the jet knots J1 and J2 move along ballistic trajectories. Assuming that the jet speed remains constant, then we can trace back to obtain the position of the jet J2 at the moments of the $\gamma$-ray and radio flares. They are in sequence: 0.041 mas (2009 $\gamma$-ray flare), 0.096 mas (2009 radio flare), 0.163 mas (2010 $\gamma$-ray flare), and  0.205 mas (2010 radio flare). The mass of the SMBH in \obj\ is $(1.84\pm0.27)\times10^8 M_\odot$ \citep{2012ApJ...755...60G}, therefore its gravitational radius is $R_g = 9\times10^{-6}$~pc. The physical size corresponding to the first $\gamma$-ray flare in 2009 is $7.6 \times 10^3 R_g$, expressed in units of gravitational radius. This size scale corresponds to the starting section of the jet, 
and it can be assumed that the collimation of the jet occurs before that distance. A smaller black hole mass of $5 \times 10^7 M_\odot$ is derived by \citet{1988A&A...198...16K}. This implies that jet collimation would occur at a longer distance, measured in units of the black hole's gravitational radius.

The second $\gamma$-ray flare and subsequent radio flare  may happen at a distance of 0.27--0.34 pc, which is consistent with the dimension of the broad line region of \obj\  \citep{1988A&A...198...16K}.
This suggests that the jet probably collides with the inner boundary of the disk wind within the broad line region, or with the clouds in the torus on a larger scale, resulting in the observed flares (see discussion in Section \ref{sec:jet-wind}).


\bibliographystyle{aasjournal}
\bibliography{references} 



\end{document}